%% file: main.tex
\pdfoutput=1
\documentclass[12pt,a4paper]{article}

\usepackage{ifthen} 
\newboolean{pdflatex}
\setboolean{pdflatex}{true} 

\newboolean{articletitles}
\setboolean{articletitles}{true} 

\newboolean{uprightparticles}
\setboolean{uprightparticles}{false} 

\def\paperauthors{LHCb collaboration} 
\def\paperasciititle{Search for the rare decay} 
\def\papertitle{Search for the decay $B^0\to\phi\mu^+\mu^-$} 
\def\paperkeywords{{High Energy Physics}, {LHCb}} 
\def\papercopyright{\the\year\ CERN for the benefit of the LHCb collaboration} 
\def\paperlicence{CC BY 4.0 licence}
\def\paperlicenceurl{https://creativecommons.org/licenses/by/4.0/}

\input{preamble}
\usepackage{longtable} 
\usepackage{multirow}
\begin{document}

\renewcommand{\thefootnote}{\fnsymbol{footnote}}
\setcounter{footnote}{1}

\input{title-LHCb-PAPER}


\renewcommand{\thefootnote}{\arabic{footnote}}
\setcounter{footnote}{0}

\cleardoublepage


\pagestyle{plain} 
\setcounter{page}{1}
\pagenumbering{arabic}


\input{1_introduction}
\input{2_detector}
\input{3_selection}
\input{4_model}
\input{5_systematic}

\input{6_conclusion}

\input{acknowledgements}




\addcontentsline{toc}{section}{References}
\bibliographystyle{LHCb}
\bibliography{main,standard,LHCb-PAPER,LHCb-CONF,LHCb-DP,LHCb-TDR,local}

\newpage
\input{Authorship_LHCb-PAPER-2021-042}

\end{document}

%% file: preamble.tex
\usepackage[top=1in, bottom=1.25in, left=1in, right=1in]{geometry}

\usepackage{microtype}
\usepackage{lineno}  
\usepackage{xspace} 
\usepackage{caption} 

\usepackage{graphicx}  
\usepackage{color}
\usepackage{colortbl}
\graphicspath{{./figs/}} 

\usepackage{amsmath} 
\usepackage{amssymb}
\usepackage{amsfonts}
\usepackage{upgreek} 

\newcommand*\patchAmsMathEnvironmentForLineno[1]{%
\expandafter\let\csname old#1\expandafter\endcsname\csname #1\endcsname
\expandafter\let\csname oldend#1\expandafter\endcsname\csname
end#1\endcsname
 \renewenvironment{#1}%
   {\linenomath\csname old#1\endcsname}%
   {\csname oldend#1\endcsname\endlinenomath}%
}
\newcommand*\patchBothAmsMathEnvironmentsForLineno[1]{%
  \patchAmsMathEnvironmentForLineno{#1}%
  \patchAmsMathEnvironmentForLineno{#1*}%
}
\AtBeginDocument{%
\patchBothAmsMathEnvironmentsForLineno{equation}%
\patchBothAmsMathEnvironmentsForLineno{align}%
\patchBothAmsMathEnvironmentsForLineno{flalign}%
\patchBothAmsMathEnvironmentsForLineno{alignat}%
\patchBothAmsMathEnvironmentsForLineno{gather}%
\patchBothAmsMathEnvironmentsForLineno{multline}%
\patchBothAmsMathEnvironmentsForLineno{eqnarray}%
}

\usepackage{hyperxmp}

\usepackage[pdftex,
            pdfauthor={\paperauthors},
            pdftitle={\paperasciititle},
            pdfkeywords={\paperkeywords},
            pdfcopyright={Copyright (C) \papercopyright},
            pdflicenseurl={\paperlicenceurl}]{hyperref}

\usepackage[colorinlistoftodos,textsize=scriptsize]{todonotes}

\usepackage[bottom,flushmargin,hang,multiple]{footmisc}

\usepackage[all]{hypcap} 

\input{lhcb-symbols-def} 

\usepackage{cite} 
\usepackage{mciteplus}

%% file: lhcb-symbols-def.tex
\usepackage{xspace} 
\usepackage{upgreek}

\def\lhcb   {\mbox{LHCb}\xspace}

\def\MagUp {\mbox{\em Mag\kern -0.05em Up}\xspace}

\ifthenelse{\boolean{uprightparticles}}%
{

 \def\Pmu         {\ensuremath{\upmu}\xspace}

 \def\Pphi        {\ensuremath{\upphi}\xspace}

 \def\Ppsi        {\ensuremath{\uppsi}\xspace}

 \def\PDelta      {\ensuremath{\Delta}\xspace}                 
 \def\PXi         {\ensuremath{\Xi}\xspace}                 
 \def\PLambda     {\ensuremath{\Lambda}\xspace}                 
 \def\PSigma      {\ensuremath{\Sigma}\xspace}                 
 \def\POmega      {\ensuremath{\Omega}\xspace}                 
 \def\PUpsilon    {\ensuremath{\Upsilon}\xspace}

 \def\PB      {\ensuremath{\mathrm{B}}\xspace}                 
                  
 \def\PD      {\ensuremath{\mathrm{D}}\xspace}

 \def\PJ      {\ensuremath{\mathrm{J}}\xspace}                 
 \def\PK      {\ensuremath{\mathrm{K}}\xspace}

 \def\Pb      {\ensuremath{\mathrm{b}}\xspace}                 
 \def\Pc      {\ensuremath{\mathrm{c}}\xspace}

 \def\Pi      {\ensuremath{\mathrm{i}}\xspace}

 \def\Ps      {\ensuremath{\mathrm{s}}\xspace}

 \def\thebaroffset{0.0em}
}
{

 \def\Pmu         {\ensuremath{\mu}\xspace}

 \def\Pphi        {\ensuremath{\phi}\xspace}

 \def\Ppsi        {\ensuremath{\psi}\xspace}                 
                  
 \mathchardef\PDelta="7101
 \mathchardef\PXi="7104
 \mathchardef\PLambda="7103
 \mathchardef\PSigma="7106
 \mathchardef\POmega="710A
 \mathchardef\PUpsilon="7107
                  
 \def\PB      {\ensuremath{B}\xspace}                 
                  
 \def\PD      {\ensuremath{D}\xspace}

 \def\PJ      {\ensuremath{J}\xspace}                 
 \def\PK      {\ensuremath{K}\xspace}

 \def\Pb      {\ensuremath{b}\xspace}                 
 \def\Pc      {\ensuremath{c}\xspace}

 \def\Pi      {\ensuremath{i}\xspace}

 \def\Ps      {\ensuremath{s}\xspace}

 \def\thebaroffset{0.18em}
}
\newcommand{\offsetoverline}[2][\thebaroffset]{\kern #1\overline{\kern -#1 #2}}%

\makeatletter
\ifcase \@ptsize \relax
  \newcommand{\miniscule}{\@setfontsize\miniscule{4}{5}}
\or
  \newcommand{\miniscule}{\@setfontsize\miniscule{5}{6}}
\or
  \newcommand{\miniscule}{\@setfontsize\miniscule{5}{6}}
\fi
\makeatother

\DeclareRobustCommand{\optbar}[1]{\shortstack{{\miniscule (\rule[.5ex]{1.25em}{.18mm})}
  \\ [-.7ex] $#1$}}

\def\mup        {{\ensuremath{\Pmu^+}}\xspace}
\def\mun        {{\ensuremath{\Pmu^-}}\xspace} 

\def\squark    {{\ensuremath{\Ps}}\xspace}

\def\cquark    {{\ensuremath{\Pc}}\xspace}

\def\bquark    {{\ensuremath{\Pb}}\xspace}

\def\kaon    {{\ensuremath{\PK}}\xspace}

\def\KorKbar {\kern \thebaroffset\optbar{\kern -\thebaroffset \PK}{}\xspace}

\def\Kp      {{\ensuremath{\kaon^+}}\xspace}
\def\Km      {{\ensuremath{\kaon^-}}\xspace}

\newcommand{\phiz}{\ensuremath{\Pphi}\xspace}

\def\D       {{\ensuremath{\PD}}\xspace}

\def\DorDbar {\kern \thebaroffset\optbar{\kern -\thebaroffset \PD}\xspace}

\def\Dp      {{\ensuremath{\D^+}}\xspace}
\def\Dm      {{\ensuremath{\D^-}}\xspace}

\def\DpDm    {\ensuremath{\Dp {\kern -0.16em \Dm}}\xspace}

\def\B       {{\ensuremath{\PB}}\xspace}

\def\BorBbar {\kern \thebaroffset\optbar{\kern -\thebaroffset \PB}\xspace}

\def\Bd      {{\ensuremath{\B^0}}\xspace}

\def\BdorBdbar {\kern \thebaroffset\optbar{\kern -\thebaroffset \Bd}\xspace}

\def\Bs      {{\ensuremath{\B^0_\squark}}\xspace}

\def\BsorBsbar {\kern \thebaroffset\optbar{\kern -\thebaroffset \Bs}\xspace}

\def\Bds     {{\ensuremath{\B_{(\squark)}^0}}\xspace}

\def\jpsi     {{\ensuremath{{\PJ\mskip -3mu/\mskip -2mu\Ppsi}}}\xspace}

\def\Y#1S{\ensuremath{\PUpsilon{(#1S)}}\xspace}

\def\Lz          {{\ensuremath{\PLambda}}\xspace}

\def\LorLbar     {\kern \thebaroffset\optbar{\kern -\thebaroffset \PLambda}\xspace}

\def\Lb           {{\ensuremath{\Lz^0_\bquark}}\xspace}

\def\to                 {\ensuremath{\rightarrow}\xspace}

\def\AT#1     {\ensuremath{A_{\mathrm{T}}^{#1}}\xspace}           

\def\C#1      {\ensuremath{\mathcal{C}_{#1}}\xspace}                       
\def\Cp#1     {\ensuremath{\mathcal{C}_{#1}^{'}}\xspace}                    
\def\Ceff#1   {\ensuremath{\mathcal{C}_{#1}^{\mathrm{(eff)}}}\xspace}        
\def\Cpeff#1  {\ensuremath{\mathcal{C}_{#1}^{'\mathrm{(eff)}}}\xspace}       
\def\Ope#1    {\ensuremath{\mathcal{O}_{#1}}\xspace}                       
\def\Opep#1   {\ensuremath{\mathcal{O}_{#1}^{'}}\xspace}                    

\newcommand{\nospaceunit}[1]{\ensuremath{\text{#1}}}       
\newcommand{\aunit}[1]{\ensuremath{\text{\,#1}}}       

\newcommand{\tev}{\aunit{Te\kern -0.1em V}\xspace}
\newcommand{\gev}{\aunit{Ge\kern -0.1em V}\xspace}
\newcommand{\mev}{\aunit{Me\kern -0.1em V}\xspace}
\newcommand{\kev}{\aunit{ke\kern -0.1em V}\xspace}
\newcommand{\ev}{\aunit{e\kern -0.1em V}\xspace}
 
\newcommand{\mevc}{\ensuremath{\aunit{Me\kern -0.1em V\!/}c}\xspace}
\newcommand{\gevc}{\ensuremath{\aunit{Ge\kern -0.1em V\!/}c}\xspace}
\newcommand{\mevcc}{\ensuremath{\aunit{Me\kern -0.1em V\!/}c^2}\xspace}
\newcommand{\gevcc}{\ensuremath{\aunit{Ge\kern -0.1em V\!/}c^2}\xspace}
\newcommand{\gevgevcc}{\ensuremath{\gev^2\!/c^2}\xspace} 
\newcommand{\gevgevcccc}{\ensuremath{\gev^2\!/c^4}\xspace} 

\def\mum  {\ensuremath{\,\upmu\nospaceunit{m}}\xspace}

\def\fb   {\ensuremath{\aunit{fb}}\xspace}
\def\invfb   {\ensuremath{\fb^{-1}}\xspace}

\newcommand{\stat}{\aunit{(stat)}\xspace}
\newcommand{\syst}{\aunit{(syst)}\xspace}

\newcommand{\chisq}{\ensuremath{\chi^2}\xspace}

\newcommand{\chisqip}{\ensuremath{\chi^2_{\text{IP}}}\xspace}

\def\gsim{{~\raise.15em\hbox{$>$}\kern-.85em
          \lower.35em\hbox{$\sim$}~}\xspace}
\def\lsim{{~\raise.15em\hbox{$<$}\kern-.85em
          \lower.35em\hbox{$\sim$}~}\xspace}

\def\pt         {\ensuremath{p_{\mathrm{T}}}\xspace}

\def\ptot       {\ensuremath{p}\xspace}

\def\evtgen     {\mbox{\textsc{EvtGen}}\xspace}

\def\geant      {\mbox{\textsc{Geant4}}\xspace}

\def\photos     {\mbox{\textsc{Photos}}\xspace}

\def\pythia     {\mbox{\textsc{Pythia}}\xspace}

\def\tell1  {TELL1\xspace}
\def\ukl1   {UKL1\xspace}



%% file: title-LHCb-PAPER.tex

\begin{titlepage}
\pagenumbering{roman}

\vspace*{-1.5cm}
\centerline{\large EUROPEAN ORGANIZATION FOR NUCLEAR RESEARCH (CERN)}
\vspace*{1.5cm}
\noindent
\begin{tabular*}{\linewidth}{lc@{\extracolsep{\fill}}r@{\extracolsep{0pt}}}
\ifthenelse{\boolean{pdflatex}}
{\vspace*{-1.5cm}\mbox{\!\!\!\includegraphics[width=.14\textwidth]{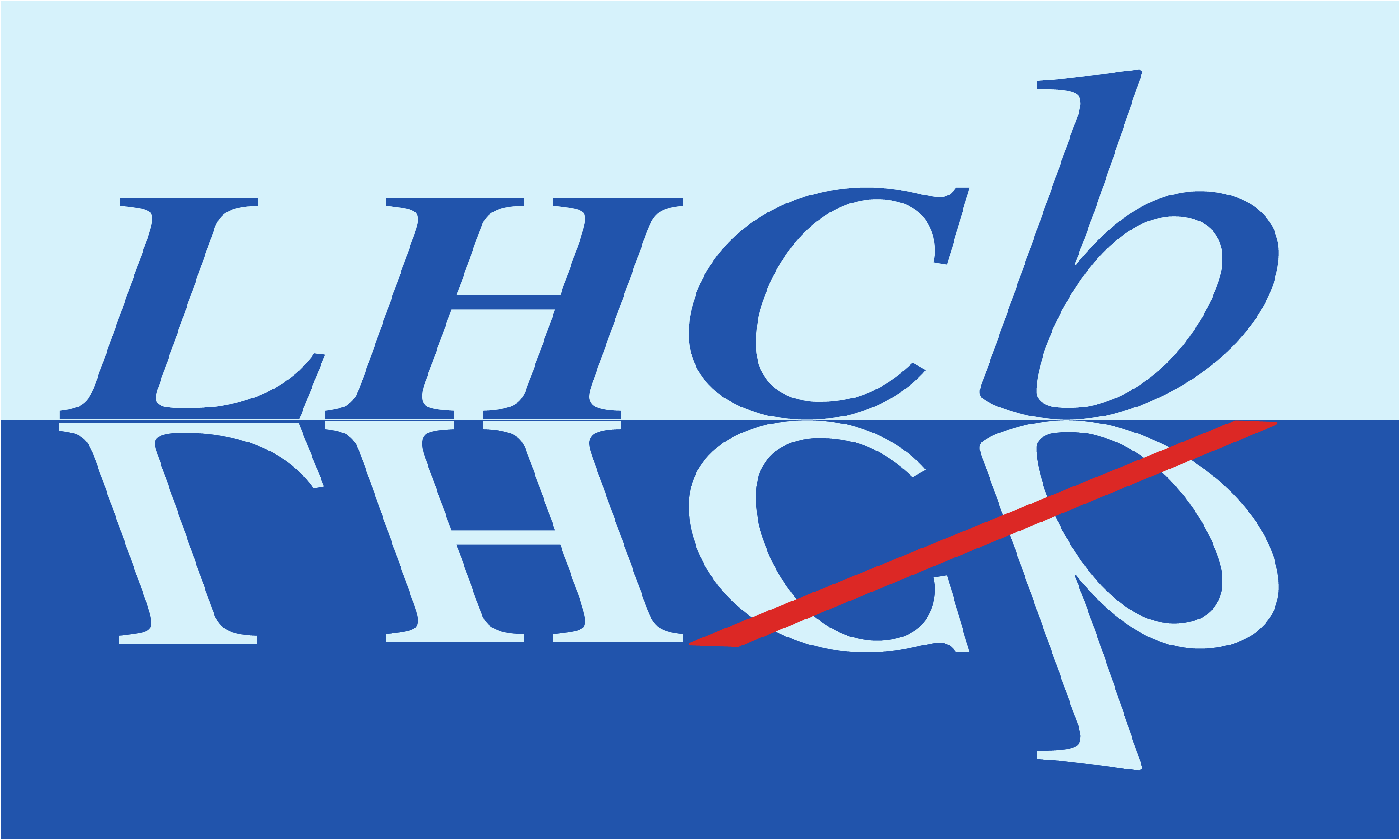}} & &}%
{\vspace*{-1.2cm}\mbox{\!\!\!\includegraphics[width=.12\textwidth]{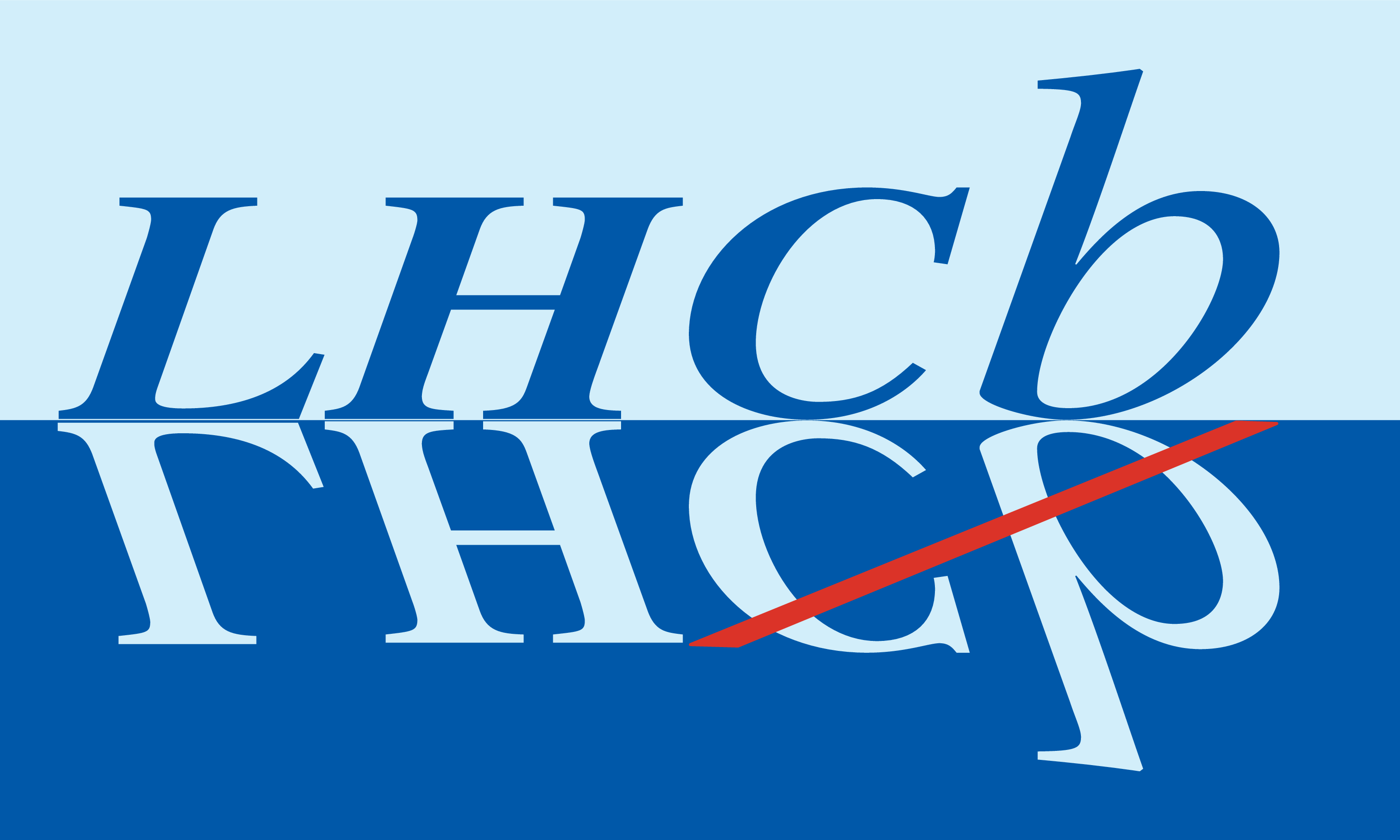}} & &}%
\\
 & & CERN-EP-2021-267 \\  
 & & LHCb-PAPER-2021-042 \\  
 & & May 11, 2022 \\
 & & \\
\end{tabular*}

\vspace*{4.0cm}

{\normalfont\bfseries\boldmath\huge
\begin{center}
  \papertitle 
\end{center}
}

\vspace*{2.0cm}

\begin{center}
\paperauthors\footnote{Authors are listed at the end of this paper.}
\end{center}

\vspace{\fill}

\begin{abstract}
  \noindent
A search for the decay $\Bd\to\phi\mu^+\mu^-$ is performed using proton-proton collisions at centre-of-mass energies of 7, 8, and 13\tev collected
by the LHCb experiment and corresponding to an integrated luminosity of 9\invfb.
No evidence for the $B^0\to \phi \mu^+ \mu^-$ decay is found and
an upper limit on the branching fraction, excluding the $\phi$ and charmonium regions in the dimuon spectrum,
of  $4.4 \times 10^{-3}$ at a 90\% credibility level, relative to that of the $B^0_s \to \phi \mu^+ \mu^-$ decay, is established. 
Using the measured $\Bs\to\phi\mu^+\mu^-$ branching fraction
and assuming a phase-space model,  the absolute branching fraction of the decay $B^0\to \phi \mu^+ \mu^-$ in the full $q^2$ range
is determined to be less than $3.2 \times 10^{-9}$ at a 90\% credibility level.
  
\end{abstract}

\vspace*{2.0cm}

\begin{center}
  Published in
  JHEP 05 (2022) 067
\end{center}

\vspace{\fill}

{\footnotesize 
\centerline{\copyright~\papercopyright. \href{\paperlicenceurl}{\paperlicence}.}}
\vspace*{2mm}

\end{titlepage}


\newpage
\setcounter{page}{2}
\mbox{~}
%
%
%
%

%% file: 1_introduction.tex
\section{Introduction}
\label{sec:Introduction}

\begin{figure}[b]
   \begin{center}
    \includegraphics*[width=0.48\linewidth]{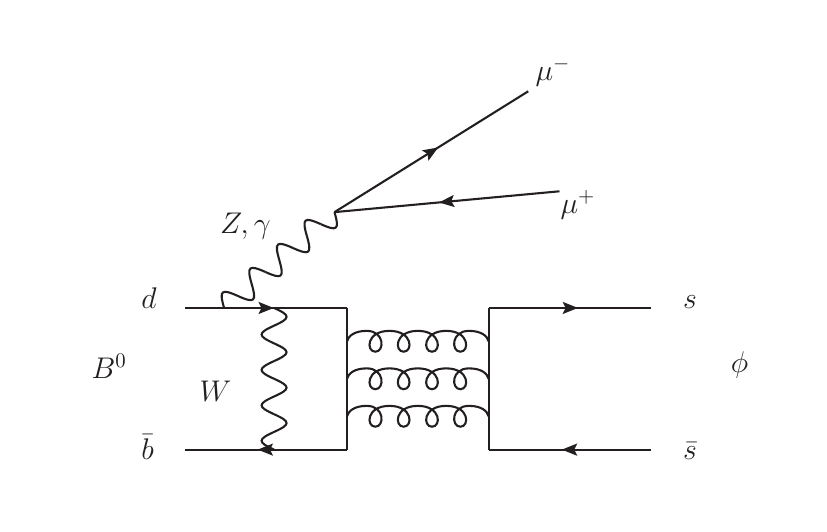}
    \put(-120,5){ (a)}
    \includegraphics*[width=0.48\linewidth]{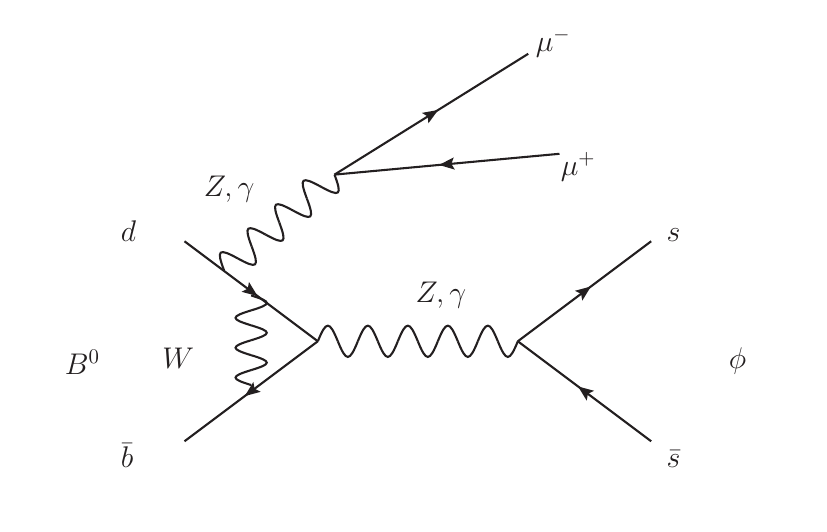}
    \put(-120,5){ (b)}\\ 
    \includegraphics*[width=0.48\linewidth]{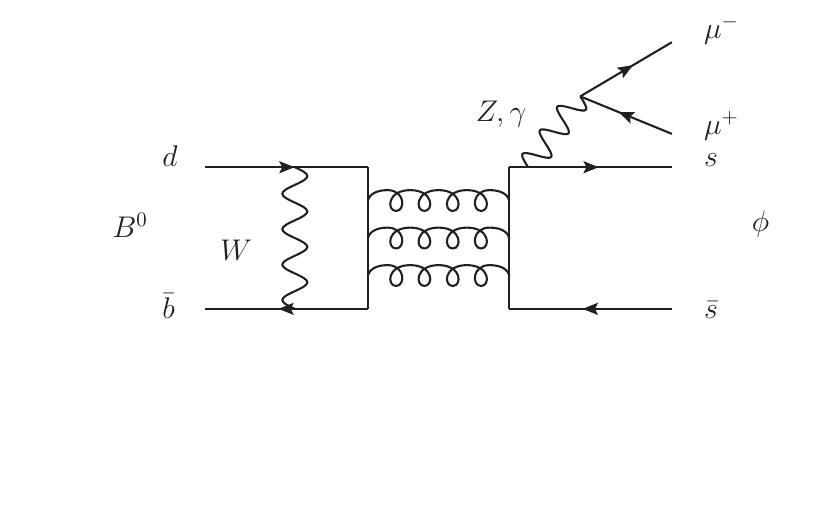}
    \hspace*{0.1in}
     \put(-125,35){ (c)}
    \includegraphics*[width=0.48\linewidth]{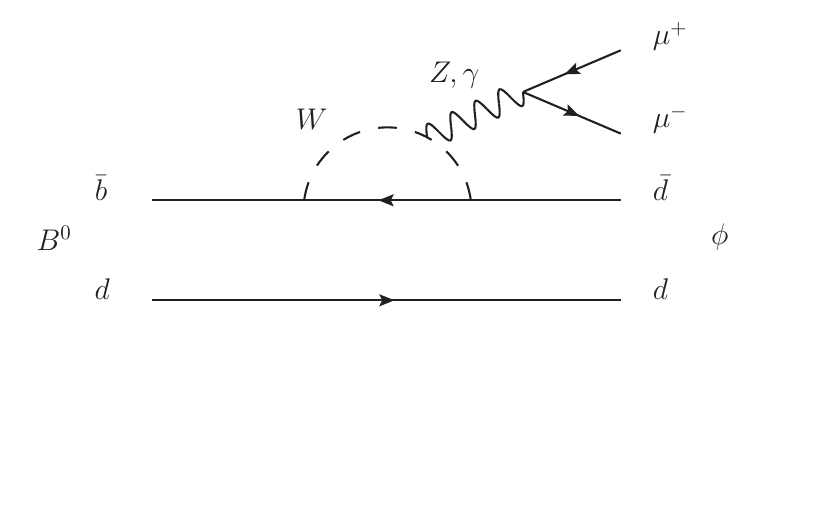}
    \put(-126,35){ (d)}
    \vspace*{-0.5cm}
    \end{center}
    \caption{Standard Model Feynman diagrams for the decay $\B^0 \rightarrow \phi\mu^+\mu^-$.
    (a), (b), (c) represent the weak annihilation contributions, 
    while (d) represents the contribution from  $\omega-\phi$ mixing. 
    }
\label{loop}
\end{figure}

The decay $\Bd \to \phi \mu^+ \mu^-$ proceeds
mainly via the color-suppressed penguin annihilation diagrams (a), (b), and (c) in Fig.~\ref{loop}, if we consider only the $s\bar{s}$ component of the $\phi$ meson.
Annihilation decays of $B$ mesons are strongly suppressed in the Standard Model (SM)
but very sensitive to physics beyond the SM.
The annihilation contribution to the $\Bd \to \phi \mu^+ \mu^-$ branching fraction is estimated
to be approximately of the order of $10^{-12}$ in the QCD factorization approach~\cite{Kuznetsova:2017ecg}. 
However, when using this decay to probe new physics, 
the contribution from the small $d\bar{d}$ component of the $\phi$ meson must be considered.
Contributions from $\omega-\phi$ mixing, Fig.~\ref{loop} (d), 
and new physics could have significant effects on this decay. 
There is no theoretical study of these effects in the literature. 
Some clues can be found in the reported studies of the decay $\Bd \to \phi \gamma$~\cite{Li:2003kz,Li:2006xe,Lu:2006nza,Hua:2010we,PhysRevD.103.076004,King:2016cxv},
which has similar quark-level transitions as the $\Bd \to \phi \mu^+ \mu^-$ decay.
The annihilation contributions to the $\Bd \to \phi \gamma$ branching fraction 
have been found to be of the order of $10^{-12}$ to $10^{-11}$~\cite{Li:2003kz,Li:2006xe,Lu:2006nza}, 
depending on the factorization techniques.
Contributions by new particles such as a $Z^{\prime}$ boson in the annihilation diagrams 
could be of the order of  $10^{-9}$ --  $10^{-8}$~\cite{Li:2003kz, Hua:2010we}, 
large enough to be observed by the LHCb detector.
A recent study with soft-collinear effective theory indicates that the contribution from $\omega-\phi$ mixing 
could be three orders of magnitude larger than the pure annihilation contribution in the SM,
increasing the branching fraction of the decay $\Bd \to \phi \gamma$
to $\mathcal{O}(10^{-9})$~\cite{PhysRevD.103.076004}.
The decay $B^0 \rightarrow \phi\gamma$ has not yet been observed,
and the current upper limit on the branching fraction is $1.0 \times 10^{-7}$ at a  $90\%$ confidence level set by the Belle collaboration~\cite{King:2016cxv}.

Assuming a dominant $\omega-\phi$ contribution~\cite{PhysRevD.103.076004} 
and scaling the $B^0 \rightarrow \rho^0 \mu^+ \mu^-$ branching fraction measured by the LHCb experiment~\cite{LHCb-PAPER-2014-063},
the $\Bd \to \phi \mu^+ \mu^-$ branching fraction is predicted to be between $10^{-11}$ and $10^{-10}$. 
The decay $\Bd \to \phi \mu^+ \mu^-$ has not yet been observed,
but may be accessible at high luminosity flavour physics experiments such as the LHCb experiment and its upgrade,
where it can be reconstructed with high efficiency.

This article presents a search for the  decay $\Bd \to \phi \mu^+ \mu^-$
performed using proton-proton ($pp$) collision  data collected with the LHCb detector,
corresponding to a total integrated luminosity of 9\invfb,
comprising 3\invfb collected at centre-of-mass energies of 7 and 8\tev during 2011 and 2012 (denoted \mbox{Run~1}) and 6\invfb collected at 13\tev from 2015 to 2018 (denoted \mbox{Run~2}). 
The search is performed in the kinematically allowed range of $q^2$,
the squared invariant mass of the dimuon system,
excluding the $\phi$ region of \mbox{0.98--1.1\gevgevcccc},
the $\jpsi$ region of \mbox{8.0--11.0\gevgevcccc},
and the $\psi(2S)$ region of \mbox{12.5--15.0\gevgevcccc}.
The decay $B_s^0 \to \phi\mu^+\mu^-$ is used as the normalization channel;
its branching fraction in the same $q^2$ regions has already been measured by the LHCb experiment~\cite{LHCb:2021zwz}.
The more copious decay $B_s^0 \to J/\psi\phi$ with $\jpsi \to \mu^+\mu^-$ has identical  final-state  products and similar kinematic distributions as $B^0_{(s)}\to \phi \mu^+\mu^-$ decays. 
A high purity sample of $B_s^0 \to J/\psi\phi$  decays is used to develop a multivariate event classifier 
and determine the mass model for nonresonant $B^0_{(s)}\to \phi \mu^+\mu^-$ decays, 
where nonresonant refers to the $\mu^+\mu^-$ pair. 

%% file: 2_detector.tex
\section{Detector and simulation}
\label{sec:Detector and simulation}

The \lhcb detector~\cite{LHCb-DP-2008-001,LHCb-DP-2014-002} is a single-arm forward
spectrometer covering the \mbox{pseudorapidity} range $2<\eta <5$,
designed for the study of particles containing \bquark or \cquark
quarks. The detector includes a high-precision tracking system
consisting of a silicon-strip vertex detector surrounding the $pp$
interaction region~\cite{LHCb-DP-2014-001}, a large-area silicon-strip detector located
upstream of a dipole magnet with a bending power of about
$4{\mathrm{\,Tm}}$, and three stations of silicon-strip detectors and straw
drift tubes~\cite{LHCb-DP-2013-003,LHCb-DP-2017-001}
placed downstream of the magnet.
The tracking system provides a measurement of the momentum, \ptot, of charged particles with
a relative uncertainty that varies from 0.5\% at low momentum to 1.0\% at 200\gevc.
The minimum distance of a track to a primary $pp$ collision vertex (PV), the impact parameter (IP),
is measured with a resolution of $(15+29/\pt)\mum$,
where \pt is the component of the momentum transverse to the beam, in\,\gevc.
Different types of charged hadrons are distinguished using information
from two ring-imaging Cherenkov detectors~\cite{LHCb-DP-2012-003}.
Photons, electrons and hadrons are identified by a calorimeter system consisting of
scintillating-pad and preshower detectors, an electromagnetic
and a hadronic calorimeter. Muons are identified by a
system composed of alternating layers of iron and multiwire
proportional chambers~\cite{LHCb-DP-2012-002}.

The online event selection is performed by a trigger~\cite{LHCb-DP-2012-004},
consisting of a hardware stage, based on information from the calorimeter and muon systems, followed by a software stage, which applies a full event reconstruction.
In the hardware stage, signal candidates are required to 
have at least one muon with \pt greater than 1 to 2\gevc or a pair of muons with the product of their \pt above 1 to 4\gevgevcc,
depending on the data-taking conditions.
The software trigger requires a two-, three- or four-track secondary vertex with a significant displacement from any PV.
At least one charged particle must have a $\pt$ greater than $1\gevc$
and be inconsistent with originating from a PV.
A multivariate algorithm~\cite{Gligorov_2013} is used for the identification of secondary vertices
consistent with the decay of a $b$ hadron.
The total trigger efficiency is 81\%, where this quantity is defined as the number of simulated signal events that pass the full selection, including the trigger, divided by the number of signal events that pass all the section criteria, except the trigger requirements.

Samples of simulated decays are used to determine the trigger, reconstruction and selection efficiencies of the signal and control channels, as well as to estimate contamination from specific background processes. In the simulation, $pp$ collisions are generated using \pythia~\cite{Sjostrand:2007gs}
with a specific \lhcb configuration~\cite{LHCb-PROC-2010-056}.
Decays of unstable particles
are described by \evtgen~\cite{Lange:2001uf}, in which final-state
radiation is generated using \photos~\cite{davidson2015photos}.
The interaction of the generated particles with the detector, and its response,
are implemented using the \geant
toolkit~\cite{Allison:2006ve, *Agostinelli:2002hh} as described in
Ref.~\cite{LHCb-PROC-2011-006}.

%% file: 3_selection.tex
\section{Candidate selection}
\label{sec:Candidate selection}

The candidates of the $\Bds \to \phi\mu^+\mu^-$ signal sample
and the $B_s^0 \to J/\psi\phi$ control sample are reconstructed
by combining a pair of oppositely charged tracks, identified as muons, and a pair of oppositely charged tracks, identified as kaons.
These tracks are required to be compatible with originating from a common vertex 
and have significant \chisqip
with respect to all primary interaction vertices,
where \chisqip is defined as the difference in the vertex-fit \chisq of a given PV reconstructed with and without the track under consideration.
The \Bds candidates must have a decay vertex significantly displaced from any PV
and be compatible with originating from one of the PVs, considered as the \Bds production vertex.
The angle between the vector connecting the
production and decay vertices and the momentum of the \Bds candidate, $\theta$, must satisfy $\cos\theta>0.999$.
The mass of the \Kp\Km\mup\mun combination is restricted to the range 
\mbox{5100--5800\mevcc} and the invariant mass of the \Kp\Km pair must be within 12\mevcc of the known \phiz mass~\cite{PDG2020}.
The $\Bds \to \phi\mu^+\mu^-$ signal candidates are selected in the $q^2$ range excluding the $\phi$ and charmonium regions, 
while the $B_s^0 \to J/\psi\phi$ candidates are required to have a $q^2$ in the $\jpsi$ region of \mbox{8.0--11.0\gevgevcccc}.

There are two major sources of peaking background.
The first consists of $B_s^0 \to J/\psi\phi$ decays 
with a muon  reconstructed as a kaon and a kaon as a muon. This background is suppressed by 
removing candidates that have a $K^{\pm}\mu^{\mp}$ mass in the $\jpsi$ region,
where a muon mass is assigned to any kaon candidate that satisfies strict criteria for muon selection.
The second peaking background is due to $\Lambda_b^0 \to p K^-\mu^+\mu^-$ decays with the proton  misidentified as a kaon.\footnote {The inclusion of charge-conjugate states is implied throughout.}
This source is suppressed by rejecting candidates in the $\Lambda_b^0$ region of the $K^+K^-\mu^+\mu^-$ mass spectrum,
where a proton mass is assigned to any kaon candidate that satisfies strict criteria for proton selection.

A boosted decision tree (BDT)~\cite{Breiman,*AdaBoost} 
classifier is employed to  reduce the combinatorial background arising from random track combinations.
The BDT input variables include the \chisqip of all final state tracks and of the \Bds candidate,
cosine of the angle $\theta$,
the fit \chisq of the \Bds decay vertex and  its displacement from the production vertex,
the \Bds transverse momentum,
the particle identification information of the final-state products,
and the multiplicity and kinematic information of tracks consistent with the \Bds decay vertex but not associated with the \Bds candidate.

Separate BDT classifiers are trained for data taken in the Run 1 and Run 2 periods. The training of each BDT 
uses a data sample enriched with  $\Bs \to \jpsi \phi$ signal candidates, 
of which each event is assigned a  weight for background subtraction using the $sPlot$ technique~\cite{Pivk:2004ty} with the mass $m(\Kp\Km \mu^+ \mu^-)$ as the discriminating variable.  The background sample used in the training 
consists of $\Kp\Km \mu^+ \mu^-$  combinations with  invariant mass of the dimuon pair outside the $\jpsi$ and $\psi(2S)$ mass regions, 
invariant mass of the dikaon pair within 50\mevcc of the known $\phi$ mass, and $m(K^+K^- \mu^+\mu^-)$ more than 200\mevcc above the known \Bs mass~\cite{PDG2020}. 
This $m(K^+K^- \mu^+\mu^-)$ sideband is chosen to avoid overlapping the mass region used in the subsequent mass fit.

The  BDT threshold is chosen to  maximize the  figure of merit for the decay $\B^0 \to \phi \mu^+ \mu^-$, defined as
\mbox{$\varepsilon/(\tfrac{a}{2} + \sqrt{B})$}~\cite{Punzi:2003bu}.
Here $\varepsilon$ is the signal efficiency of the BDT requirement, which is estimated using a data sample of  $B_s^0 \to J/\psi\phi$ candidates independent of the $B_s^0 \to J/\psi\phi$  sample used for BDT training.
The background yield, $B$, in the $B^0$ signal mass window of [5249, 5309]\mevcc
is estimated via interpolation between the lower sideband of [5170, 5249]\mevcc
and upper sideband of [5309, 5570]\mevcc.
The targeted significance, $a$,  is set to $3$.
The same  BDT requirement is used for the
selection of $B^0 \to \phi \mu^+ \mu^-$, $B_s^0 \to \phi \mu^+ \mu^-$, and $B_s^0 \to J/\psi\phi$ decays. 
The distributions of the BDT input variables and the efficiency of the BDT requirement  are  found to be similar in the three channels according to the  simulation. This ensures that the BDT classifier trained and optimized using the $B_s^0 \to J/\psi\phi$ control sample is also optimal for the $\Bds \to \phi \mu^+ \mu^-$ channels.
The BDT classifier rejects about 99\% of the combinatorial background, while keeping about 80\% of the signal and control channnel candidates.

The reconstruction and selection efficiencies needed 
for the branching fraction calculation are determined using simulated samples of $\Bd \to \phi \mu^+\mu^-$ and $\Bs \to \phi \mu^+\mu^-$ decays, which are generated using a phase-space model and an amplitude model with inputs from Ref.~\cite{PhysRevD.71.014029}.
The simulation is corrected for imperfect modeling  of the particle identification performance, the track multiplicity, the distributions of  transverse momentum,  and  vertex fit $\chi^2$  of the \Bds mesons, using the $\Bs \to \jpsi \phi$ control sample from the data.
The ratio of the average efficiencies for $B^0 \to \phi \mu^+ \mu^-$ and $B_s^0 \to \phi \mu^+ \mu^-$ decays with $q^2$ outside the $\phi$ and charmonium regions is evaluated to be 
$\varepsilon(B^0\to \phi\mu^+\mu^-)/\varepsilon(B_s^0 \to \phi\mu^+\mu^-)= 0.999 \pm 0.009  $ for Run 1 and $0.969 \pm 0.007  $ for Run 2, respectively. 
Here, the uncertainties are due to limited size of the simulation samples. 

%% file: 4_model.tex
\section{Mass fits}
\label{sec:Mass fit}

\begin{figure}[h]
    \begin{center}
        \includegraphics*[width=0.49\textwidth]{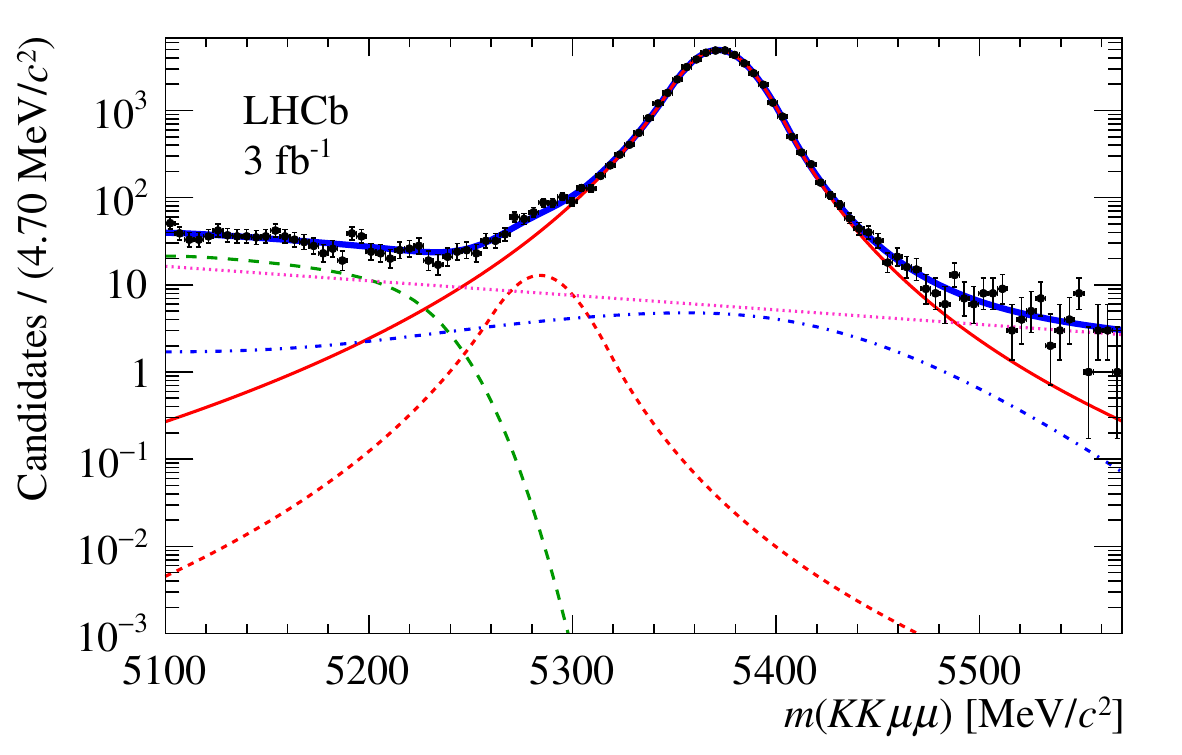}
        \includegraphics*[width=0.49\textwidth]{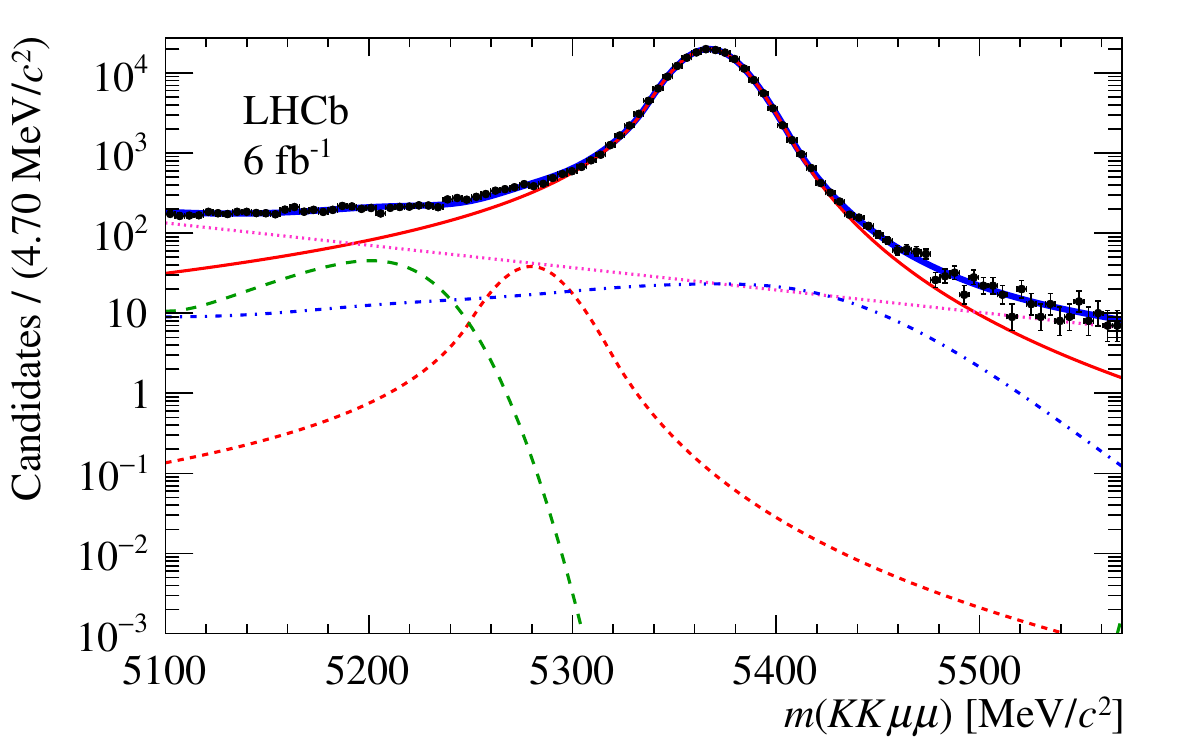}
    \end{center}
    \caption{The $\Kp\Km\mu^+\mu^-$ mass distributions of selected $B_s^0 \to J/\psi\phi$  candidates in (left) \mbox{Run 1} and (right) Run 2 data, with the fit projections  overlaid.
      The red solid line is $\Bs \to \jpsi\phi$ signal, 
      the red dashed line is $\Bd \to \jpsi \Kp\Km$ signal,
      the green dashed line is the partially reconstructed background component,
      the violet dotted line is the combinatorial background component,
      and the blue dash-dot line is the \mbox{$\Lb\to \jpsi{p}K^-$} background component.
    }
    \label{resonant}
\end{figure}

The branching fraction of the  nonresonant decay $\Bd \to \phi \mu^+\mu^-$ relative to that of the decay $\Bs \to \phi \mu^+\mu^-$ is estimated from a fit to the  $\Kp\Km \mu^+\mu^-$ mass distribution in a range  that contains both the $\Bd$ and $\Bs$ signal peaks.
The signal mass shape
of the $\Bds \to \phi \mu^+\mu^-$ decays
is partially determined using the $B_s^0 \to J/\psi\phi$  control sample. 
The $\Kp\Km\mu^+\mu^-$ mass distribution of $B_s^0 \to J/\psi\phi$ candidates in the range $\mbox{5100--5570\mevcc}$ is shown in Fig.~\ref{resonant}.
An unbinned maximum-likelihood fit is performed to this distribution, separately for Run 1 and Run 2 data.
The $B_s^0 \to J/\psi\phi$  candidates are reconstructed and selected in the same way as the nonresonant candidates, with  no $\jpsi$ mass constraint  applied.
The probability density function (PDF) for this fit is the sum of a
$B_s^0 \to J/\psi\phi$ component, a $\Bd\to \jpsi \Kp\Km$ component, and three background components. 
The $B_s^0 \to J/\psi\phi$ component is described by a double-sided Hypatia function~\cite{Santos:2013gra}, with tail parameters  obtained from the fit. 
The $\Bd\to \jpsi \Kp\Km$ component has the same shape as that of the $B_s^0 \to J/\psi\phi$ decay, and the difference of their mean values is constrained to the difference of the  known $\Bd$ and $\Bs$ masses~\cite{PDG2020}. 
The $\Bd\to \jpsi \Kp\Km$ yield is fixed to the estimate of $119 \pm 19$ for Run 1 ($362 \pm 51$ for Run 2) obtained {\it a priori} from another mass fit where the invariant mass of the $B_s^0 \to J/\psi\phi$ candidates is computed with a $\jpsi$ mass constraint applied on the dimuon pair. 

\begin{figure}[ht]
    \begin{center}
        \includegraphics*[width=0.49\textwidth]{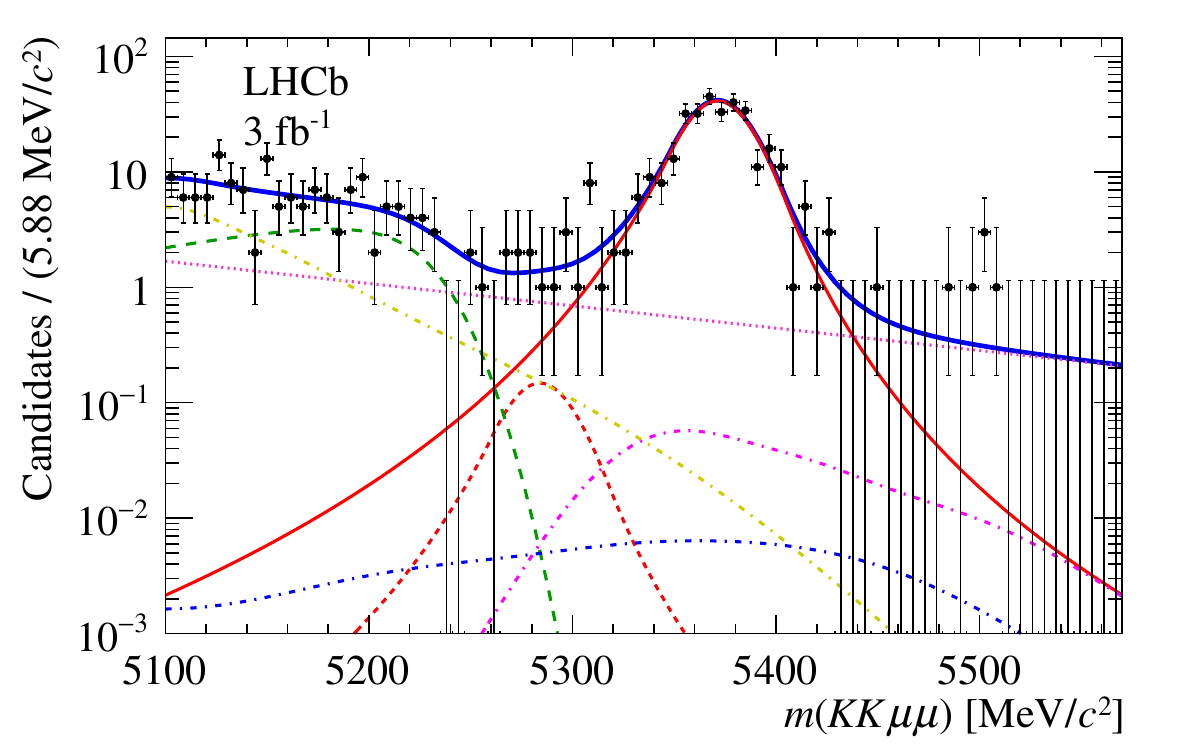}
        \includegraphics*[width=0.49\textwidth]{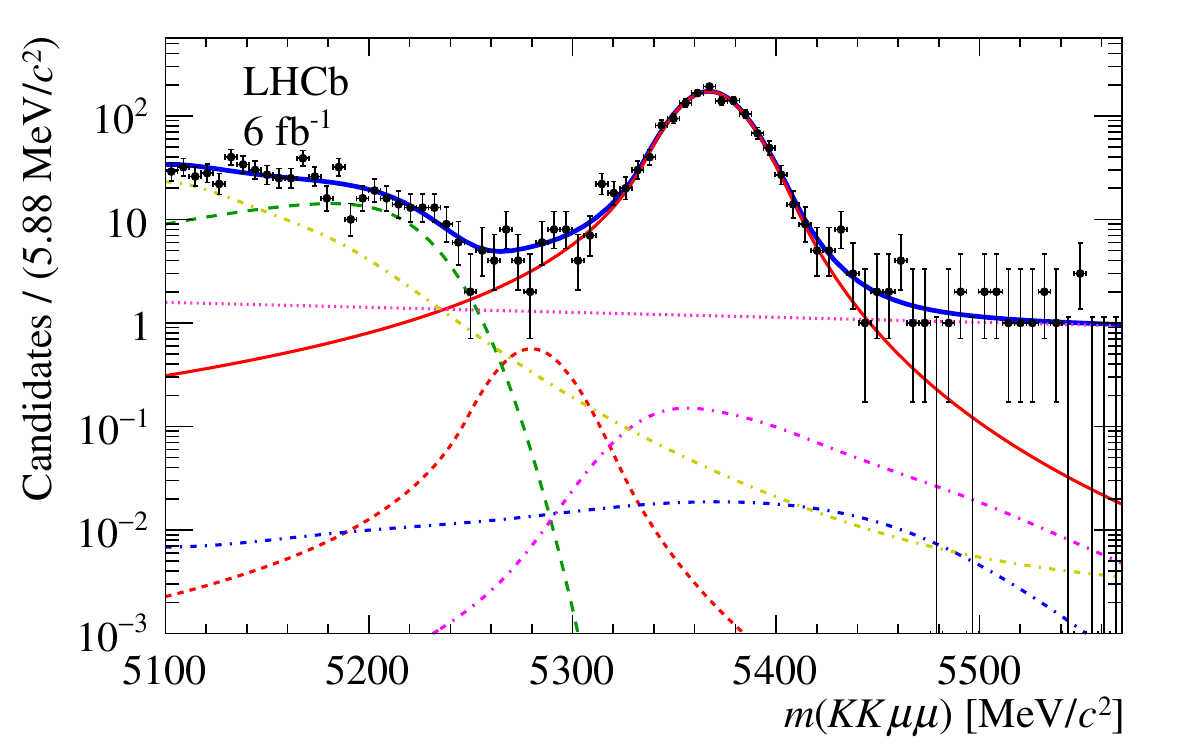}
    \end{center}
    \caption{The $\Kp\Km\mu^+\mu^-$ mass distributions of selected nonresonant $\Bds \to \phi\mu^+\mu^-$ candidates in (left) Run 1 and (right) Run 2 data.
      The red solid line is $\Bs \to \phi\mu^+\mu^-$ signal, 
      the red dashed line is $\Bd \to \phi\mu^+\mu^-$ signal,
      the green dashed line is the partially reconstructed background component,
      the violet dotted line is the combinatorial background component,
      the blue dash-dot line is the \mbox{$\Lb\to pK^-\mu^+\mu^-$} background component,
      the violet dash-dot line is the \mbox{$B^0\to K^{*0}(\to K^+\pi^-)\mu^+\mu^-$} background component,
      and the orange dash-dot line is the \mbox{$\Bs \to D_s^-(\to \phi\mu^-\bar{\nu})\mu^+\nu$} background component.
    }
    \label{nonresonant}
\end{figure}

The combinatorial background for $\Bs\to \jpsi \phi$
 is described by an exponential function.
The residual  background from $\Lb\to J/\psi{p}K^-$ decays 
passing the dedicated veto is described by a template 
obtained from simulation. The $\Lb\to J/\psi{p}K^-$ yield is fixed to the estimate of 
 $253 \pm 53$ for Run 1 ($1251 \pm 172$ for Run 2), which is obtained {\it a priori} by changing the mass hypothesis of one kaon to a proton and 
fitting the $pK^{-}\mu^+\mu^-$ mass distribution, following the procedure described in Refs.~\cite{LHCb-PAPER-2019-013, LHCb-PAPER-2020-033}. 
The partially reconstructed background mainly arises from $B$-meson decays to final states with a $\pi^0$, and is modelled by an Argus function~\cite{Verkerke:2003ir} convolved with a Gaussian resolution function with a width  equal to that of the signal Hypatia function. 
The endpoint of the Argus function is fixed to the mean of the \Bs mass peak minus the $\pi^0$ mass~\cite{PDG2020}. 
The fit projections of the $\Kp\Km\mu^+\mu^-$ mass distributions of selected $B_s^0 \to J/\psi\phi$ candidates are shown in Fig.~\ref{resonant} and are in good agreement with data.

A simultaneous unbinned maximum-likelihood fit is performed to the $\Kp\Km\mu^+\mu^-$ mass distributions of selected $\Bds\to \phi \mu^+\mu^-$ candidates, shown in Fig.~\ref{nonresonant}, in the Run 1 and Run 2 data samples.
The fit range is $\mbox{5100--5570\mevcc}$.
The  fit model  detailed below  keeps the same form for \mbox{Run 1} and Run 2, while the fit parameters can take different values for the two periods except for the parameter of interest,  
the branching fraction ratio 
in the $q^2$ range excluding the $\phi$ and charmonium regions,
\begin{equation}
\mathcal{R}=\frac{\mathcal{B}(\Bd\to\phi\mu^+\mu^-)}{\mathcal{B}(\Bs\to\phi\mu^+\mu^-)} \;,
\end{equation}
which is required to be common  for Run 1 and Run 2.
The fit PDF includes the $\Bd \to \phi \mu^+\mu^-$ and  $\Bs \to \phi \mu^+\mu^-$ components; a combinatorial background component;
several additional background components from specific $B$-meson decays: \mbox{$B^0\to K^{*0}(\to K^+\pi^-)\mu^+\mu^-$}, \mbox{$\Lb\to pK^-\mu^+\mu^-$}, \mbox{$\Bs \to D_s^-(\to \phi\mu^-\bar{\nu})\mu^+\nu$}; and an inclusive partially reconstructed background component.

As in the $B_s^0 \to J/\psi\phi$ case, the $\Bs \to \phi \mu^+\mu^-$ component is described by a  double-sided Hypatia function and its tail parameters are fixed to the values obtained in the  $B_s^0 \to J/\psi\phi$ fit. 
The width, mean, and yield ($N_{\Bs}$)  are allowed to vary in the fit. The $\Bd \to \phi \mu^+\mu^-$ component is described by the same double-sided Hypatia function as for \Bs decays
shifted by the  difference of the known \Bd and \Bs masses. The branching fraction ratio $\mathcal{R}$ is included as a free fit parameter. 
The $\Bd\to\phi \mu^+\mu^-$ yield ($N_{\Bd}$) is expressed in terms of  $N_{\Bs}$ and  $\mathcal{R}$ according to   
\begin{equation}
     N_{B^0}    =   \frac{\mathcal{R}}{f_s/f_d} \times 
    \frac{\varepsilon(B^0\to \phi\mu^+\mu^-)}{\varepsilon(B_s^0 \to \phi\mu^+\mu^-)}
    \times N_{B_s^0}\;.
    \label{Nbsb0}
\end{equation}
Here $\varepsilon(B^0\to \phi\mu^+\mu^-)/\varepsilon(B_s^0 \to \phi\mu^+\mu^-)$ is the efficiency ratio given in Section~\ref{sec:Candidate selection}, and $f_s/f_d$ is the ratio of the production fractions of $B_s^0$ and $B^0$ mesons in the LHCb detector acceptance in $pp$ collisions, which has been measured to  be 0.2390 $\pm$ 0.0076 at 7\tev, 0.2385 $\pm$ 0.0075 at 8\tev, and 0.2539 $\pm$ 0.0079 at 13\tev~\cite{LHCb:2021qbv}.    
The factors $f_s/f_d$ and 
${\varepsilon(B^0 \to \phi\mu^+\mu^-)} / {\varepsilon(B_s^0 \to \phi\mu^+\mu^-)}$
are fixed to their central values in the baseline fit, and their uncertainties are taken into account 
in the evaluation of the systematic uncertainties of the $\mathcal{R}$ measurement. 

As in the $B_s^0 \to J/\psi\phi$ case,
the combinatorial background for $\Bds\to \phi\mu^+\mu^-$ is described by an exponential function, the inclusive partially reconstructed background is  modelled by an Argus function convolved with a Gaussian resolution function, and the Argus endpoint is set to the mean of the $B_s^0$ mass peak minus  the $\pi^0$ mass. 

Three sources of specific physics background are accounted for in the \mbox{$\Bds\to \phi \mu^+\mu^-$} mass fit, \mbox{$B^0\to K^{*0}(\to K^+\pi^-)\mu^+\mu^-$} decays with the pion misidentified as a kaon,
residual \mbox{$\Lb\to pK^-\mu^+\mu^-$} decays with the proton misidentified as a kaon and \mbox{$\Bs \to D_s^-(\to \phi\mu^-\bar{\nu})\mu^+\nu$} decays with the two neutrinos undetected. 
Their mass models are implemented as templates  obtained from corrected simulation.
The yields are determined relative to the $\Bs\to \phi \mu^+ \mu^-$ yield,
using the known branching fractions and the efficiencies relative to that of $\Bs\to \phi \mu^+ \mu^-$ given in Table~\ref{tab:effratiotab}.
The obtained yields are  $N_{\Bd\to K^{*0}\mu^+\mu^-}=1.21 \pm 0.23$, $N_{\Lb\to pK^-\mu^+\mu^-}=0.29 \pm 0.12$, 
and $N_{\Bs\to D_s^- \mu^+\nu}=52 \pm 17$ for Run 1 ($2.87 \pm 0.51$, $0.87 \pm 0.35$, and $240 \pm 77$ for \mbox{Run 2}).
The central values of these  yields are used in the baseline fit and their uncertainties are considered as sources of
systematic uncertainties for the $\mathcal{R}$ estimate. 

\begin{table}[t]
    \begin{center}
    \caption{Efficiencies of  background decay processes  relative to that of the decay $B_s^0 \to \phi \mu^+ \mu^-$ evaluated using simulated samples. 
     The uncertainties are due to limited sizes of these  samples.}
        \begin{tabular}{l|c|c}
            \hline
            \multirow{2}{*}{Process}  &  \multicolumn{2}{c}{$\varepsilon_{\rm bkg}/\varepsilon_{B_s^0}\; [\times10^{-3}] $}  \\ \cline{2-3}
                     & Run 1 & Run 2 \\ \hline
            $B^0 \to K^{*0} \mu^+\mu^-$  & $0.671 \pm 0.041$ & $0.344 \pm 0.018$ \\ 
            $\Lb \to p K^- \mu^+ \mu^-$  & $0.717 \pm 0.033$ & $0.469 \pm 0.016$ \\ 
            $B_s^0 \to D_s^- \mu^+ \nu$ & $0.298 \pm 0.015$ & $0.299 \pm 0.008$ \\ \hline
        \end{tabular}
            \label{tab:effratiotab}
    \end{center}
\end{table}

The  $\Bs\to \phi \mu^+ \mu^-$ signal yields are  $302\pm 19 $ for Run 1 and $1389 \pm 41$ for Run 2. The fit projections are shown in Fig.~\ref{nonresonant},
and there is no visible $\Bd \to \phi \mu^+ \mu^-$ signal contribution.
Thus the upper limits on its relative and absolute branching fractions are calculated in Section~\ref{sec:results}.

%% file: 5_systematic.tex
\section{Systematic uncertainties}
\label{sec:Systematic uncertainties}

Due to the identical decay products and similar event topology of the $\Bd\to \phi\mu^+\mu^-$ and $\Bs\to \phi \mu^+\mu^-$ decays,  
systematic uncertainties associated with the evaluation of the efficiency
cancel in the branching fraction ratio $\mathcal{R}$. 
The remaining systematic uncertainties, including additive ones associated with the yield estimation
and multiplicative ones propagated from the
scaling factors involved in the calculation of $\mathcal{R}$,  are  summarized in Table~\ref{tab:systab} and discussed below.

\begin{table}[b]
    \begin{center}
    \caption{Systematic uncertainties on the measurement of $\mathcal{R}$ for additive and multiplicative sources. }
        \begin{tabular}{l c}
            \hline
            Additive uncertainties & Value $[\times10^{-3}]$ \\ \hline
            Fit bias  & 0.09  \\ 
            Signal model  & 0.39 \\ 
            Partial background  &  0.15 \\ 
            Yield of $B^0 \to J/\psi K^+ K^-$   & 0.09 \\ 
            Yield of $\Lb \to J/\psi p K^- $  & 0.07 \\ 
            Yield of $B^0 \to K^{*0} \mu^+\mu^-$  & 0.01 \\ 
            Yield of $\Lb \to p K^- \mu^+ \mu^-$   & 0.03 \\ 
            Yield of $B_s^0 \to D_s^- \mu^+ \nu$  & 0.27  \\
            Shape of $B^0 \to K^{*0} \mu^+\mu^-$  & 0.01 \\
            Shape of $\Lb \to p K^- \mu^+ \mu^-$  & 0.00 \\
            Shape of $B_s^0 \to D_s^- \mu^+ \nu$  & 0.13  \\
            \textbf{Total} & \textbf{0.54} \\ \hline
            Multiplicative uncertainties & Value $[\%]$ \\ \hline
            ${f_s}/{f_d}$ & 3.1 \\ 
            $\varepsilon_{B^0} / \varepsilon_{B_s^0}$ & 0.8 \\ 
            \textbf{Total} & \textbf{3.2} \\ \hline            
        \end{tabular}
     \label{tab:systab}
    \end{center}
\end{table}

The dominant systematic uncertainty is associated with modelling the mass shapes of the signals. 
This effect has been studied by fitting the data using an alternative model, generating a large number of samples according to the obtained new model, and fitting each pseudoexperiments 
with both the baseline and alternative model.
The mean change in $\mathcal{R}$ is assigned as a systematic uncertainty.  For $\Bds\to\phi \mu^+\mu^-$ decays,
replacing the double-sided Hypatia function with the sum of two double-sided Crystal Ball  functions 
leads to an uncertainty of $0.39\times 10^{-3}$ on  $\mathcal{R}$.
For the inclusive partially reconstructed  background, changing the resolution model from a Gaussian to a Hypatia function causes an uncertainty of $0.15\times 10^{-3}$. 

Another major contribution to the systematic uncertainty is associated with the specific background from $B_s^0 \to D_s^-(\to \phi \mu^- \bar{\nu}) \mu^+ \nu$ decays with missing neutrinos, which lies under the inclusive partially reconstructed  background in the $\Kp\Km\mu^+\mu^-$ mass spectrum.
The shape of this background is described by a template obtained from simulation.
The uncertainty due to the finite size of the simulated sample is evaluated using a bootstrapping technique~\cite{efron1979}. 
A large number of new samples of the same size as the original simulation sample are formed by randomly cloning events from the original sample. 
The standard deviation on the results of $\mathcal{R}$ obtained using the new samples is taken as a  systematic uncertainty, which is estimated to be $0.13\times 10^{-3}$. In the baseline fit, 
the yield of the $B_s^0 \to D_s^- \mu^+ \nu$ background is fixed to the central value of the estimate given in  Section~\ref{sec:Mass fit}. 
Changing this yield by $\pm 1$ standard deviations and repeating the $\Bds\to \phi \mu^+\mu^-$ mass fit, the maximum change of  $\mathcal{R}$ is $0.27\times 10^{-3}$, which is assigned as a systematic uncertainty. 

The systematic uncertainties associated with other specific background components in the $\Bds\to \phi \mu^+\mu^-$ sample are also studied and found to be small. Changing the fixed yield of the $\Bd\to\jpsi \Kp\Km$ ($\Lb\to \jpsi p\Km$) component in the
$\Bs\to \jpsi \phi$ fit by $\pm 1$ standard deviations leads to  a systematic uncertainty  of  $0.09\times 10^{-3}$ ($0.07\times 10^{-3}$) on $\mathcal{R}$. 
The average bias in $\mathcal{R}$
due to the maximum likelihood fit procedure is evaluated to be $0.09\times 10^{-3}$  using pseudoexperiments. Summing the contributions discussed above in quadrature leads to a total additive systematic uncertainty of $\sigma_{\rm add}=0.54\times 10^{-3}$.

As can be seen in Eq.~\ref{Nbsb0}, the estimate of  $\mathcal{R}$ is proportional to the production fraction ratio $f_s / f_d$ and  
the efficiency ratio ${\varepsilon_{\Bs}}/{\varepsilon_{\Bd}}$.
In the baseline fit, these  scaling factors are fixed to their central values obtained {\it a priori}.  
The  relative uncertainties of the  luminosity-averaged values of $f_s/f_d$ and ${\varepsilon_{\Bd}}/{\varepsilon_{\Bs}}$ 
are  $3.1\%$ and $0.8\%$, respectively, which are propagated  to $\mathcal{R}$ as multiplicative systematic uncertainties.  
The combined multiplicative uncertainty on $\mathcal{R}$ is $k=3.2\%$.
The total systematic uncertainty on $\mathcal{R}$  can be written as 
\begin{equation}
    \sigma(\mathcal{R}) = \sqrt{\sigma_{\rm add}^2 +(k\times{\mathcal{R}})^2}\;.
    \label{sys_tot}
\end{equation}

\section{Results}
\label{sec:results}

Since no significant signal of the decay $\Bd\to\phi \mu^+\mu^-$ is observed,  
an upper limit on the branching fraction ratio $\mathcal{R}$ is determined using the  profile likelihood method~\cite{Cowan:2010js,Schott:2012zb}.
The profile likelihood ratio as a function of $\mathcal{R}$, denoted  $\lambda_0(\mathcal{R})$, 
is defined as the ratio of the maximum likelihood value for a given value of the parameter of interest, $\mathcal{R}$,
to the global maximum likelihood value.
In order to incorporate the systematic uncertainties, a smeared profile likelihood ratio function is defined as 
\begin{equation}
 \lambda (\mathcal{R}) =  \lambda_0({\mathcal{R}^{\prime }})  \otimes G({\mathcal{R} - \mathcal{R}^{\prime}}  ; 0, \sigma(  {\mathcal{R}^{\prime}}  )) \;, 
 \label{lambda_smear}
\end{equation}
where $\lambda_0(\mathcal{R}^{\prime})$ is  convolved with a Gaussian function, which has a zero mean and a width equal to the total systematic uncertainty given in Eq.~\ref{sys_tot}.

\begin{figure}[t]
\begin{center}
\includegraphics*[width=0.75\textwidth]{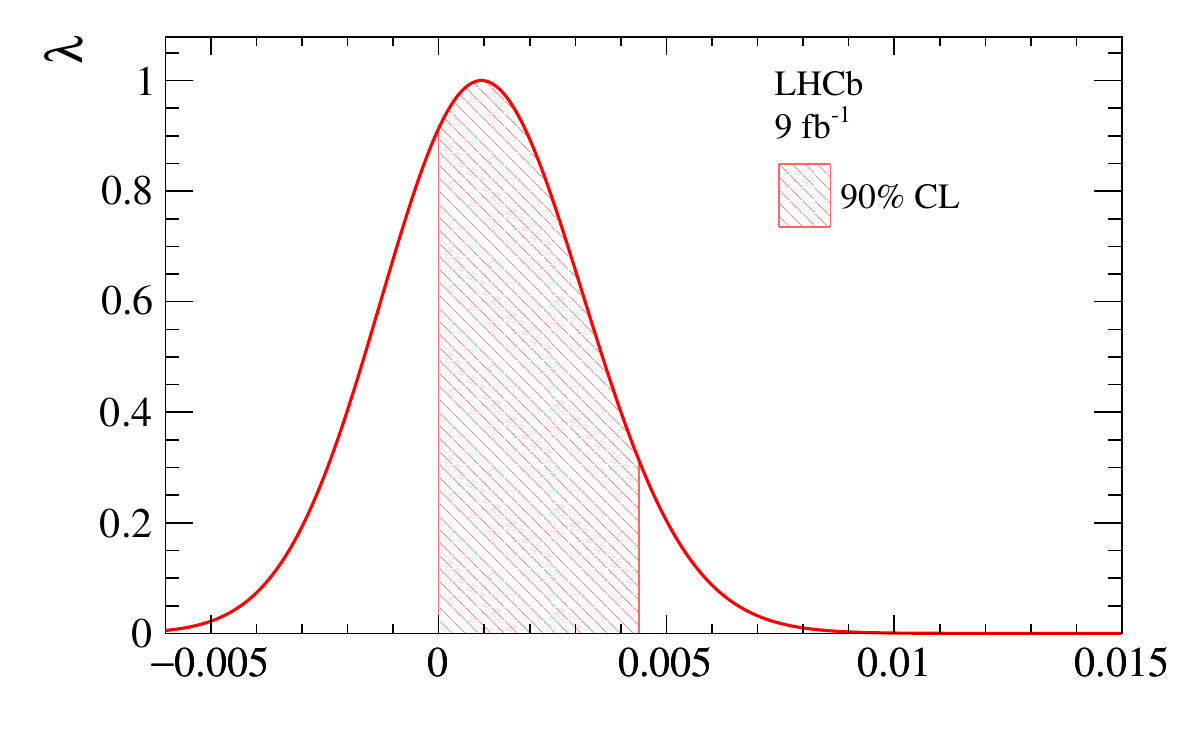}
\put( -30,5){\large{$\mathcal{R}$}}
\end{center}
\caption{Smeared profile likelihood ratio curve from the simultaneous fit to the Run 1 and \mbox{Run 2} data samples.  The red shaded area indicates the $90\%$ crediblity interval of $\mathcal{R}$. 
\label{unblind_uplimt_smear}
} 
\end{figure}

Figure~\ref{unblind_uplimt_smear} shows the smeared likelihood function $\lambda(\mathcal{R})$  obtained from the simultaneous fit to the Run 1 and Run 2 data samples,
where the shaded area starting at $\mathcal{R} = 0$
defines a $90\%$ credibility interval obtained using  a prior function that is uniform in the physical region $\mathcal{R} >0$.
The right boundary of this interval gives the upper limit 
on $\mathcal{R}$ 
\begin{equation}
    \mathcal{R} < 4.4\times 10^{-3}\;\mbox{\rm at a 90\% crediblity level (CL)}\;. \notag
    \label{limit_R}
\end{equation}

The limit on  $\mathcal{R}$  can be converted into a limit on the branching fraction $\mathcal{B}(B^0 \to \phi\mu^+\mu^-)$ using a previous  measurement of $\mathcal{B}(B_s^0 \to \phi\mu^+\mu^-)$  in the same $q^2$ intervals. 
The LHCb collaboration reported $\mathcal{B}(B_s^0 \to \phi\mu^+\mu^-) =(8.14 \pm 0.21 \stat \pm 0.16 \syst \pm 0.03\,(\rm{extrap})$ $\pm 0.39\,(B_s^0 \to \jpsi \phi)) \times 10^{-7}$~\cite{LHCb:2021zwz} 
in the full $q^2$ range without the resonant vetoes, where the third uncertainty is associated with the extrapolation  used to recover the vetoed $\phi$ and charmonium regions  in the $q^2$ spectrum. 
Using the extrapolation factor of $F^s_e = (65.47 \pm 0.27)\%$ given in Ref.~\cite{LHCb:2021zwz}, the branching fraction  excluding the $\phi$ and charmonium regions  is $\mathcal{B}(B_s^0 \to \phi\mu^+\mu^-) =(5.33 \pm 0.14 \stat \pm 0.10 \syst \pm 0.25\, (B_s^0 \to \jpsi \phi)) \times 10^{-7}$. 
Among these uncertainties, the contributions from  the $\Bs \to \phi \mu^+\mu^-$ yield  and $f_s/f_d$ ratio  are almost completely anticorrelated with the corresponding uncertainties on   $\mathcal{R}$.
Taking this correlation into account, the net uncertainty propagated from $\mathcal{B}(\Bs \to \phi\mu^+\mu^-)$ to $\mathcal{B}(\Bd \to \phi\mu^+\mu^-)$  is  found to be negligible.
A limit  on $\mathcal{B}(\Bd \to \phi\mu^+\mu^-)$ in the $q^2$ range excluding the $\phi$ and charmonium regions is set to be  $2.3\times 10^{-9} $ at a $90\%$ CL. 

The fraction of $\Bd\to \phi \mu^+\mu^-$ decays within the considered $q^2$ regions is calculated to be $F^d_e = (73.2\pm 0.1 )\%$ with a phase-space decay model.
Using this fraction,  the limits on the  total branching fractions in the full $q^2$ range is determined to be
\begin{equation}
    \mathcal{B}(\Bd \to \phi\mu^+\mu^-) < 3.2\times 10^{-9}\; \mbox{\rm at a 90\% CL}\;. \notag
\end{equation}
The observed limit on   $\mathcal{B}(\Bd \to \phi\mu^+\mu^-) $ is consistent with  the
expected limit, which  is evaluated to be  $3.1\times 10^{-9}$ at a 90\% CL using pseudoexperiments generated under the assumption of zero $\Bd \to \phi \mu^+\mu^-$  signal.  
Alternative models are used to check the dependency of the result on the  $\Bd \to \phi \mu^+\mu^-$  decay model. 
The phase-space model is replaced by a model that has the same  $q^2$ and angular distributions as in  $\Bs \to \phi \mu^+\mu^-$ decays or a model that has the same $q^2$ distribution as in $\Bs \to \phi \mu^+\mu^-$ decays but a flat angular distribution.
The evaluated upper limits on $\mathcal{B}(\Bd \to \phi\mu^+\mu^-) $ increase by less than $5\%$ and $15\%$ in the $q^2$ range, excluding the $\phi$ and charmonium resonances, and in the full $q^2$ range, respectively.

%% file: 6_conclusion.tex
\section{Conclusion}
\label{sec:Conclusion}
This article presents the first search for the decay $B^0 \to \phi\mu^+\mu^-$,  performed using $pp$ collision data  at centre-of-mass energies of 7, 8, and 13\tev
collected by the LHCb experiment, corresponding to an integrated luminosity of  9\invfb. 
No statistically significant excess of the decay $B^0 \to \phi\mu^+\mu^-$ above the background is observed. 
An upper limit on its branching fraction excluding the $\phi$ and charmonium regions in the dimuon spectrum
relative to that of the decay $B_s^0 \to \phi\mu^+\mu^-$ 
is determined to be  $4.4\times 10^{-3}$ at a 90\% CL. 
Assuming a phase-space decay model for the decay $B^0\to \phi \mu^+ \mu^-$ and using the LHCb measurement of  $\mathcal{B}(\Bs \to \phi \mu^+ \mu^-)$, 
an upper limit on $\mathcal{B}(\Bd \to \phi\mu^+\mu^-)$ in the full $q^2$ range is set to be $3.2 \times 10^{-9}$ at a 90$\%$ CL,
which is compatible with the SM prediction.

%% file: acknowledgements.tex
\section*{Acknowledgements}
%
%
\noindent We express our gratitude to our colleagues in the CERN
accelerator departments for the excellent performance of the LHC. We
thank the technical and administrative staff at the LHCb
institutes.
We acknowledge support from CERN and from the national agencies:
CAPES, CNPq, FAPERJ and FINEP (Brazil); 
MOST and NSFC (China); 
CNRS/IN2P3 (France); 
BMBF, DFG and MPG (Germany); 
INFN (Italy); 
NWO (Netherlands); 
MNiSW and NCN (Poland); 
MEN/IFA (Romania); 
MSHE (Russia); 
MICINN (Spain); 
SNSF and SER (Switzerland); 
NASU (Ukraine); 
STFC (United Kingdom); 
DOE NP and NSF (USA).
We acknowledge the computing resources that are provided by CERN, IN2P3
(France), KIT and DESY (Germany), INFN (Italy), SURF (Netherlands),
PIC (Spain), GridPP (United Kingdom), RRCKI and Yandex
LLC (Russia), CSCS (Switzerland), IFIN-HH (Romania), CBPF (Brazil),
PL-GRID (Poland) and NERSC (USA).
We are indebted to the communities behind the multiple open-source
software packages on which we depend.
Individual groups or members have received support from
ARC and ARDC (Australia);
AvH Foundation (Germany);
EPLANET, Marie Sk\l{}odowska-Curie Actions and ERC (European Union);
A*MIDEX, ANR, IPhU and Labex P2IO, and R\'{e}gion Auvergne-Rh\^{o}ne-Alpes (France);
Key Research Program of Frontier Sciences of CAS, CAS PIFI, CAS CCEPP, 
Fundamental Research Funds for the Central Universities, 
and Sci. \& Tech. Program of Guangzhou (China);
RFBR, RSF and Yandex LLC (Russia);
GVA, XuntaGal and GENCAT (Spain);
the Leverhulme Trust, the Royal Society
 and UKRI (United Kingdom).

%% file: Authorship_LHCb-PAPER-2021-042.tex
\centerline
{\large\bf LHCb collaboration}
\begin
{flushleft}
\small
R.~Aaij$^{32}$,
A.S.W.~Abdelmotteleb$^{56}$,
C.~Abell{\'a}n~Beteta$^{50}$,
F.~Abudin{\'e}n$^{56}$,
T.~Ackernley$^{60}$,
B.~Adeva$^{46}$,
M.~Adinolfi$^{54}$,
H.~Afsharnia$^{9}$,
C.~Agapopoulou$^{13}$,
C.A.~Aidala$^{87}$
% ...

{\footnotesize \it

% ...
$^{9}$Universit{\'e} Clermont Auvergne, CNRS/IN2P3, LPC, Clermont-Ferrand, France\\
% ...
$^{13}$LPNHE, Sorbonne Universit{\'e}, Paris Diderot Sorbonne Paris Cit{\'e}, CNRS/IN2P3, Paris, France\\
% ...
$^{32}$Nikhef National Institute for Subatomic Physics, Amsterdam, Netherlands\\
% ...
$^{46}$Instituto Galego de F{\'\i}sica de Altas Enerx{\'\i}as (IGFAE), Universidade de Santiago de Compostela, Santiago de Compostela, Spain\\
% ...
$^{50}$Physik-Institut, Universit{\"a}t Z{\"u}rich, Z{\"u}rich, Switzerland\\
% ...
$^{54}$H.H. Wills Physics Laboratory, University of Bristol, Bristol, United Kingdom\\
% ...
$^{56}$Department of Physics, University of Warwick, Coventry, United Kingdom\\
% ...
$^{60}$Oliver Lodge Laboratory, University of Liverpool, Liverpool, United Kingdom\\
% ...
$^{68}$Syracuse University, Syracuse, NY, United States\\
% ...
$^{87}$University of Michigan, Ann Arbor, United States, associated to $^{68}$\\
\bigskip
% ...
}
\end{flushleft}

%% file: main.bbl
\ifx\mcitethebibliography\mciteundefinedmacro
\PackageError{LHCb.bst}{mciteplus.sty has not been loaded}
{This bibstyle requires the use of the mciteplus package.}\fi
\providecommand{\href}[2]{#2}
\begin{mcitethebibliography}{10}
\mciteSetBstSublistMode{n}
\mciteSetBstMaxWidthForm{subitem}{\alph{mcitesubitemcount})}
\mciteSetBstSublistLabelBeginEnd{\mcitemaxwidthsubitemform\space}
{\relax}{\relax}

\bibitem{Kuznetsova:2017ecg}
A.~Kuznetsova and A.~Parkhomenko,
  \ifthenelse{\boolean{articletitles}}{\emph{{Annihilation type semileptonic
  $B$ meson decays}},
  }{}\href{https://doi.org/10.1051/epjconf/201715803003}{EPJ Web Conf.\
  \textbf{158} (2017) 03003}\relax
\mciteBstWouldAddEndPuncttrue
\mciteSetBstMidEndSepPunct{\mcitedefaultmidpunct}
{\mcitedefaultendpunct}{\mcitedefaultseppunct}\relax
\EndOfBibitem
\bibitem{Li:2003kz}
X.-Q. Li, G.-R. Lu, R.-M. Wang, and Y.~D. Yang,
  \ifthenelse{\boolean{articletitles}}{\emph{{The rare $\bar{B}^0_d \rightarrow
  \phi \gamma$ decays in Standard Model and as a probe of R parity violation}},
  }{}\href{https://doi.org/10.1140/epjc/s2004-01878-1}{Eur.\ Phys.\ J.\
  \textbf{C36} (2004) 97},
  \href{http://arxiv.org/abs/hep-ph/0305283}{{\normalfont\ttfamily
  arXiv:hep-ph/0305283}}\relax
\mciteBstWouldAddEndPuncttrue
\mciteSetBstMidEndSepPunct{\mcitedefaultmidpunct}
{\mcitedefaultendpunct}{\mcitedefaultseppunct}\relax
\EndOfBibitem
\bibitem{Li:2006xe}
Y.~Li and C.-D. Lu, \ifthenelse{\boolean{articletitles}}{\emph{{Annihilation
  type radiative decays of B meson in perturbative QCD approach}},
  }{}\href{https://doi.org/10.1103/PhysRevD.74.097502}{Phys.\ Rev.\
  \textbf{D74} (2006) 097502},
  \href{http://arxiv.org/abs/hep-ph/0605220}{{\normalfont\ttfamily
  arXiv:hep-ph/0605220}}\relax
\mciteBstWouldAddEndPuncttrue
\mciteSetBstMidEndSepPunct{\mcitedefaultmidpunct}
{\mcitedefaultendpunct}{\mcitedefaultseppunct}\relax
\EndOfBibitem
\bibitem{Lu:2006nza}
C.-D. Lu, Y.-L. Shen, and W.~Wang,
  \ifthenelse{\boolean{articletitles}}{\emph{{Role of electromagnetic dipole
  operator in the electroweak penguin dominated vector meson decays of B
  meson}}, }{}\href{https://doi.org/10.1088/0256-307X/23/10/017}{Chin.\ Phys.\
  Lett.\  \textbf{23} (2006) 2684},
  \href{http://arxiv.org/abs/hep-ph/0606092}{{\normalfont\ttfamily
  arXiv:hep-ph/0606092}}\relax
\mciteBstWouldAddEndPuncttrue
\mciteSetBstMidEndSepPunct{\mcitedefaultmidpunct}
{\mcitedefaultendpunct}{\mcitedefaultseppunct}\relax
\EndOfBibitem
\bibitem{Hua:2010we}
J.~Hua, C.~S. Kim, and Y.~Li,
  \ifthenelse{\boolean{articletitles}}{\emph{{Annihilation type charmless
  radiative decays of B meson in Non-universal $Z^\prime$ model}},
  }{}\href{https://doi.org/10.1140/epjc/s10052-010-1395-2}{Eur.\ Phys.\ J.\
  \textbf{C69} (2010) 139},
  \href{http://arxiv.org/abs/1002.2531}{{\normalfont\ttfamily
  arXiv:1002.2531}}\relax
\mciteBstWouldAddEndPuncttrue
\mciteSetBstMidEndSepPunct{\mcitedefaultmidpunct}
{\mcitedefaultendpunct}{\mcitedefaultseppunct}\relax
\EndOfBibitem
\bibitem{PhysRevD.103.076004}
H.~Deng {\em et~al.}, \ifthenelse{\boolean{articletitles}}{\emph{Study on pure
  annihilation type $b\ensuremath{\rightarrow}v\ensuremath{\gamma}$ decays},
  }{}\href{https://doi.org/10.1103/PhysRevD.103.076004}{Phys.\ Rev.\
  \textbf{D103} (2021) 076004}\relax
\mciteBstWouldAddEndPuncttrue
\mciteSetBstMidEndSepPunct{\mcitedefaultmidpunct}
{\mcitedefaultendpunct}{\mcitedefaultseppunct}\relax
\EndOfBibitem
\bibitem{King:2016cxv}
Belle collaboration, Z.~King {\em et~al.},
  \ifthenelse{\boolean{articletitles}}{\emph{{Search for the decay $B^0
  \rightarrow \phi \gamma$}},
  }{}\href{https://doi.org/10.1103/PhysRevD.93.111101}{Phys.\ Rev.\
  \textbf{D93} (2016) 111101},
  \href{http://arxiv.org/abs/1603.06546}{{\normalfont\ttfamily
  arXiv:1603.06546}}\relax
\mciteBstWouldAddEndPuncttrue
\mciteSetBstMidEndSepPunct{\mcitedefaultmidpunct}
{\mcitedefaultendpunct}{\mcitedefaultseppunct}\relax
\EndOfBibitem
\bibitem{LHCb-PAPER-2014-063}
LHCb collaboration, R.~Aaij {\em et~al.},
  \ifthenelse{\boolean{articletitles}}{\emph{{Study of the rare \Bs and \Bz
  decays into the $\pip\pim\mumu$ final state}},
  }{}\href{https://doi.org/10.1016/j.physletb.2015.02.010}{Phys.\ Lett.\
  \textbf{B743} (2015) 46},
  \href{http://arxiv.org/abs/1412.6433}{{\normalfont\ttfamily
  arXiv:1412.6433}}\relax
\mciteBstWouldAddEndPuncttrue
\mciteSetBstMidEndSepPunct{\mcitedefaultmidpunct}
{\mcitedefaultendpunct}{\mcitedefaultseppunct}\relax
\EndOfBibitem
\bibitem{LHCb:2021zwz}
LHCb collaboration, R.~Aaij {\em et~al.},
  \ifthenelse{\boolean{articletitles}}{\emph{{Branching fraction measurements
  of the rare $B^0_s\rightarrow\phi\mu^+\mu^-$ and $B^0_s\rightarrow
  f_2^\prime(1525)\mu^+\mu^-$ decays}},
  }{}\href{https://doi.org/10.1103/PhysRevLett.127.151801}{Phys.\ Rev.\ Lett.\
  \textbf{127} (2021) 151801},
  \href{http://arxiv.org/abs/2105.14007}{{\normalfont\ttfamily
  arXiv:2105.14007}}\relax
\mciteBstWouldAddEndPuncttrue
\mciteSetBstMidEndSepPunct{\mcitedefaultmidpunct}
{\mcitedefaultendpunct}{\mcitedefaultseppunct}\relax
\EndOfBibitem
\bibitem{LHCb-DP-2008-001}
LHCb collaboration, A.~A. Alves~Jr.\ {\em et~al.},
  \ifthenelse{\boolean{articletitles}}{\emph{{The \lhcb detector at the LHC}},
  }{}\href{https://doi.org/10.1088/1748-0221/3/08/S08005}{JINST \textbf{3}
  (2008) S08005}\relax
\mciteBstWouldAddEndPuncttrue
\mciteSetBstMidEndSepPunct{\mcitedefaultmidpunct}
{\mcitedefaultendpunct}{\mcitedefaultseppunct}\relax
\EndOfBibitem
\bibitem{LHCb-DP-2014-002}
LHCb collaboration, R.~Aaij {\em et~al.},
  \ifthenelse{\boolean{articletitles}}{\emph{{LHCb detector performance}},
  }{}\href{https://doi.org/10.1142/S0217751X15300227}{Int.\ J.\ Mod.\ Phys.\
  \textbf{A30} (2015) 1530022},
  \href{http://arxiv.org/abs/1412.6352}{{\normalfont\ttfamily
  arXiv:1412.6352}}\relax
\mciteBstWouldAddEndPuncttrue
\mciteSetBstMidEndSepPunct{\mcitedefaultmidpunct}
{\mcitedefaultendpunct}{\mcitedefaultseppunct}\relax
\EndOfBibitem
\bibitem{LHCb-DP-2014-001}
R.~Aaij {\em et~al.}, \ifthenelse{\boolean{articletitles}}{\emph{{Performance
  of the LHCb Vertex Locator}},
  }{}\href{https://doi.org/10.1088/1748-0221/9/09/P09007}{JINST \textbf{9}
  (2014) P09007}, \href{http://arxiv.org/abs/1405.7808}{{\normalfont\ttfamily
  arXiv:1405.7808}}\relax
\mciteBstWouldAddEndPuncttrue
\mciteSetBstMidEndSepPunct{\mcitedefaultmidpunct}
{\mcitedefaultendpunct}{\mcitedefaultseppunct}\relax
\EndOfBibitem
\bibitem{LHCb-DP-2013-003}
R.~Arink {\em et~al.}, \ifthenelse{\boolean{articletitles}}{\emph{{Performance
  of the LHCb Outer Tracker}},
  }{}\href{https://doi.org/10.1088/1748-0221/9/01/P01002}{JINST \textbf{9}
  (2014) P01002}, \href{http://arxiv.org/abs/1311.3893}{{\normalfont\ttfamily
  arXiv:1311.3893}}\relax
\mciteBstWouldAddEndPuncttrue
\mciteSetBstMidEndSepPunct{\mcitedefaultmidpunct}
{\mcitedefaultendpunct}{\mcitedefaultseppunct}\relax
\EndOfBibitem
\bibitem{LHCb-DP-2017-001}
P.~d'Argent {\em et~al.}, \ifthenelse{\boolean{articletitles}}{\emph{{Improved
  performance of the LHCb Outer Tracker in LHC Run 2}},
  }{}\href{https://doi.org/10.1088/1748-0221/12/11/P11016}{JINST \textbf{12}
  (2017) P11016}, \href{http://arxiv.org/abs/1708.00819}{{\normalfont\ttfamily
  arXiv:1708.00819}}\relax
\mciteBstWouldAddEndPuncttrue
\mciteSetBstMidEndSepPunct{\mcitedefaultmidpunct}
{\mcitedefaultendpunct}{\mcitedefaultseppunct}\relax
\EndOfBibitem
\bibitem{LHCb-DP-2012-003}
M.~Adinolfi {\em et~al.},
  \ifthenelse{\boolean{articletitles}}{\emph{{Performance of the \lhcb RICH
  detector at the LHC}},
  }{}\href{https://doi.org/10.1140/epjc/s10052-013-2431-9}{Eur.\ Phys.\ J.\
  \textbf{C73} (2013) 2431},
  \href{http://arxiv.org/abs/1211.6759}{{\normalfont\ttfamily
  arXiv:1211.6759}}\relax
\mciteBstWouldAddEndPuncttrue
\mciteSetBstMidEndSepPunct{\mcitedefaultmidpunct}
{\mcitedefaultendpunct}{\mcitedefaultseppunct}\relax
\EndOfBibitem
\bibitem{LHCb-DP-2012-002}
A.~A. Alves~Jr.\ {\em et~al.},
  \ifthenelse{\boolean{articletitles}}{\emph{{Performance of the LHCb muon
  system}}, }{}\href{https://doi.org/10.1088/1748-0221/8/02/P02022}{JINST
  \textbf{8} (2013) P02022},
  \href{http://arxiv.org/abs/1211.1346}{{\normalfont\ttfamily
  arXiv:1211.1346}}\relax
\mciteBstWouldAddEndPuncttrue
\mciteSetBstMidEndSepPunct{\mcitedefaultmidpunct}
{\mcitedefaultendpunct}{\mcitedefaultseppunct}\relax
\EndOfBibitem
\bibitem{LHCb-DP-2012-004}
R.~Aaij {\em et~al.}, \ifthenelse{\boolean{articletitles}}{\emph{{The \lhcb
  trigger and its performance in 2011}},
  }{}\href{https://doi.org/10.1088/1748-0221/8/04/P04022}{JINST \textbf{8}
  (2013) P04022}, \href{http://arxiv.org/abs/1211.3055}{{\normalfont\ttfamily
  arXiv:1211.3055}}\relax
\mciteBstWouldAddEndPuncttrue
\mciteSetBstMidEndSepPunct{\mcitedefaultmidpunct}
{\mcitedefaultendpunct}{\mcitedefaultseppunct}\relax
\EndOfBibitem
\bibitem{Gligorov_2013}
V.~V. Gligorov and M.~Williams,
  \ifthenelse{\boolean{articletitles}}{\emph{Efficient, reliable and fast
  high-level triggering using a bonsai boosted decision tree},
  }{}\href{https://doi.org/10.1088/1748-0221/8/02/p02013}{Journal of
  Instrumentation \textbf{8} (2013) P02013}\relax
\mciteBstWouldAddEndPuncttrue
\mciteSetBstMidEndSepPunct{\mcitedefaultmidpunct}
{\mcitedefaultendpunct}{\mcitedefaultseppunct}\relax
\EndOfBibitem
\bibitem{Sjostrand:2007gs}
T.~Sj\"{o}strand, S.~Mrenna, and P.~Skands,
  \ifthenelse{\boolean{articletitles}}{\emph{{A brief introduction to PYTHIA
  8.1}}, }{}\href{https://doi.org/10.1016/j.cpc.2008.01.036}{Comput.\ Phys.\
  Commun.\  \textbf{178} (2008) 852},
  \href{http://arxiv.org/abs/0710.3820}{{\normalfont\ttfamily
  arXiv:0710.3820}}\relax
\mciteBstWouldAddEndPuncttrue
\mciteSetBstMidEndSepPunct{\mcitedefaultmidpunct}
{\mcitedefaultendpunct}{\mcitedefaultseppunct}\relax
\EndOfBibitem
\bibitem{LHCb-PROC-2010-056}
I.~Belyaev {\em et~al.}, \ifthenelse{\boolean{articletitles}}{\emph{{Handling
  of the generation of primary events in Gauss, the LHCb simulation
  framework}}, }{}\href{https://doi.org/10.1088/1742-6596/331/3/032047}{J.\
  Phys.\ Conf.\ Ser.\  \textbf{331} (2011) 032047}\relax
\mciteBstWouldAddEndPuncttrue
\mciteSetBstMidEndSepPunct{\mcitedefaultmidpunct}
{\mcitedefaultendpunct}{\mcitedefaultseppunct}\relax
\EndOfBibitem
\bibitem{Lange:2001uf}
D.~J. Lange, \ifthenelse{\boolean{articletitles}}{\emph{{The EvtGen particle
  decay simulation package}},
  }{}\href{https://doi.org/10.1016/S0168-9002(01)00089-4}{Nucl.\ Instrum.\
  Meth.\  \textbf{A462} (2001) 152}\relax
\mciteBstWouldAddEndPuncttrue
\mciteSetBstMidEndSepPunct{\mcitedefaultmidpunct}
{\mcitedefaultendpunct}{\mcitedefaultseppunct}\relax
\EndOfBibitem
\bibitem{davidson2015photos}
N.~Davidson, T.~Przedzinski, and Z.~Was,
  \ifthenelse{\boolean{articletitles}}{\emph{{PHOTOS interface in C++:
  Technical and physics documentation}},
  }{}\href{https://doi.org/https://doi.org/10.1016/j.cpc.2015.09.013}{Comp.\
  Phys.\ Comm.\  \textbf{199} (2016) 86},
  \href{http://arxiv.org/abs/1011.0937}{{\normalfont\ttfamily
  arXiv:1011.0937}}\relax
\mciteBstWouldAddEndPuncttrue
\mciteSetBstMidEndSepPunct{\mcitedefaultmidpunct}
{\mcitedefaultendpunct}{\mcitedefaultseppunct}\relax
\EndOfBibitem
\bibitem{Allison:2006ve}
Geant4 collaboration, J.~Allison {\em et~al.},
  \ifthenelse{\boolean{articletitles}}{\emph{{Geant4 developments and
  applications}}, }{}\href{https://doi.org/10.1109/TNS.2006.869826}{IEEE
  Trans.\ Nucl.\ Sci.\  \textbf{53} (2006) 270}\relax
\mciteBstWouldAddEndPuncttrue
\mciteSetBstMidEndSepPunct{\mcitedefaultmidpunct}
{\mcitedefaultendpunct}{\mcitedefaultseppunct}\relax
\EndOfBibitem
\bibitem{Agostinelli:2002hh}
Geant4 collaboration, S.~Agostinelli {\em et~al.},
  \ifthenelse{\boolean{articletitles}}{\emph{{Geant4: A simulation toolkit}},
  }{}\href{https://doi.org/10.1016/S0168-9002(03)01368-8}{Nucl.\ Instrum.\
  Meth.\  \textbf{A506} (2003) 250}\relax
\mciteBstWouldAddEndPuncttrue
\mciteSetBstMidEndSepPunct{\mcitedefaultmidpunct}
{\mcitedefaultendpunct}{\mcitedefaultseppunct}\relax
\EndOfBibitem
\bibitem{LHCb-PROC-2011-006}
M.~Clemencic {\em et~al.}, \ifthenelse{\boolean{articletitles}}{\emph{{The
  \lhcb simulation application, Gauss: Design, evolution and experience}},
  }{}\href{https://doi.org/10.1088/1742-6596/331/3/032023}{J.\ Phys.\ Conf.\
  Ser.\  \textbf{331} (2011) 032023}\relax
\mciteBstWouldAddEndPuncttrue
\mciteSetBstMidEndSepPunct{\mcitedefaultmidpunct}
{\mcitedefaultendpunct}{\mcitedefaultseppunct}\relax
\EndOfBibitem
\bibitem{PDG2020}
Particle Data Group, P.~A. Zyla {\em et~al.},
  \ifthenelse{\boolean{articletitles}}{\emph{{\href{http://pdg.lbl.gov/}{Review
  of particle physics}}}, }{}\href{https://doi.org/10.1093/ptep/ptaa104}{Prog.\
  Theor.\ Exp.\ Phys.\  \textbf{2020} (2020) 083C01}\relax
\mciteBstWouldAddEndPuncttrue
\mciteSetBstMidEndSepPunct{\mcitedefaultmidpunct}
{\mcitedefaultendpunct}{\mcitedefaultseppunct}\relax
\EndOfBibitem
\bibitem{Breiman}
L.~Breiman, J.~H. Friedman, R.~A. Olshen, and C.~J. Stone, {\em Classification
  and regression trees}, Wadsworth international group, Belmont, California,
  USA, 1984\relax
\mciteBstWouldAddEndPuncttrue
\mciteSetBstMidEndSepPunct{\mcitedefaultmidpunct}
{\mcitedefaultendpunct}{\mcitedefaultseppunct}\relax
\EndOfBibitem
\bibitem{AdaBoost}
Y.~Freund and R.~E. Schapire, \ifthenelse{\boolean{articletitles}}{\emph{A
  decision-theoretic generalization of on-line learning and an application to
  boosting}, }{}\href{https://doi.org/10.1006/jcss.1997.1504}{J.\ Comput.\
  Syst.\ Sci.\  \textbf{55} (1997) 119}\relax
\mciteBstWouldAddEndPuncttrue
\mciteSetBstMidEndSepPunct{\mcitedefaultmidpunct}
{\mcitedefaultendpunct}{\mcitedefaultseppunct}\relax
\EndOfBibitem
\bibitem{Pivk:2004ty}
M.~Pivk and F.~R. Le~Diberder,
  \ifthenelse{\boolean{articletitles}}{\emph{{sPlot: A statistical tool to
  unfold data distributions}},
  }{}\href{https://doi.org/10.1016/j.nima.2005.08.106}{Nucl.\ Instrum.\ Meth.\
  \textbf{A555} (2005) 356},
  \href{http://arxiv.org/abs/physics/0402083}{{\normalfont\ttfamily
  arXiv:physics/0402083}}\relax
\mciteBstWouldAddEndPuncttrue
\mciteSetBstMidEndSepPunct{\mcitedefaultmidpunct}
{\mcitedefaultendpunct}{\mcitedefaultseppunct}\relax
\EndOfBibitem
\bibitem{Punzi:2003bu}
G.~Punzi, \ifthenelse{\boolean{articletitles}}{\emph{{Sensitivity of searches
  for new signals and its optimization}}, }{}eConf \textbf{C030908} (2003)
  MODT002, \href{http://arxiv.org/abs/physics/0308063}{{\normalfont\ttfamily
  arXiv:physics/0308063}}\relax
\mciteBstWouldAddEndPuncttrue
\mciteSetBstMidEndSepPunct{\mcitedefaultmidpunct}
{\mcitedefaultendpunct}{\mcitedefaultseppunct}\relax
\EndOfBibitem
\bibitem{PhysRevD.71.014029}
P.~Ball and R.~Zwicky,
  \ifthenelse{\boolean{articletitles}}{\emph{${B}_{d,s}\ensuremath{\rightarrow}\ensuremath{\rho},\ensuremath{\omega},{K}^{*},\ensuremath{\phi}$
  decay form factors from light-cone sum rules reexamined},
  }{}\href{https://doi.org/10.1103/PhysRevD.71.014029}{Phys.\ Rev.\
  \textbf{D71} (2005) 014029}\relax
\mciteBstWouldAddEndPuncttrue
\mciteSetBstMidEndSepPunct{\mcitedefaultmidpunct}
{\mcitedefaultendpunct}{\mcitedefaultseppunct}\relax
\EndOfBibitem
\bibitem{Santos:2013gra}
D.~Mart{\'\i}nez~Santos and F.~Dupertuis,
  \ifthenelse{\boolean{articletitles}}{\emph{{Mass distributions marginalized
  over per-event errors}},
  }{}\href{https://doi.org/10.1016/j.nima.2014.06.081}{Nucl.\ Instrum.\ Meth.\
  \textbf{A764} (2014) 150},
  \href{http://arxiv.org/abs/1312.5000}{{\normalfont\ttfamily
  arXiv:1312.5000}}\relax
\mciteBstWouldAddEndPuncttrue
\mciteSetBstMidEndSepPunct{\mcitedefaultmidpunct}
{\mcitedefaultendpunct}{\mcitedefaultseppunct}\relax
\EndOfBibitem
\bibitem{LHCb-PAPER-2019-013}
LHCb collaboration, R.~Aaij {\em et~al.},
  \ifthenelse{\boolean{articletitles}}{\emph{{Updated measurement of
  time-dependent \CP-violating observables in \mbox{\decay{\Bs}{\jpsi \Kp\Km}}
  decays}}, }{}\href{https://doi.org/10.1140/epjc/s10052-019-7159-8}{Eur.\
  Phys.\ J.\  \textbf{C79} (2019) 706}, Erratum
  \href{https://doi.org/10.1140/epjc/s10052-020-7875-0}{ibid.\   \textbf{C80}
  (2020) 601}, \href{http://arxiv.org/abs/1906.08356}{{\normalfont\ttfamily
  arXiv:1906.08356}}\relax
\mciteBstWouldAddEndPuncttrue
\mciteSetBstMidEndSepPunct{\mcitedefaultmidpunct}
{\mcitedefaultendpunct}{\mcitedefaultseppunct}\relax
\EndOfBibitem
\bibitem{LHCb-PAPER-2020-033}
LHCb collaboration, R.~Aaij {\em et~al.},
  \ifthenelse{\boolean{articletitles}}{\emph{{Search for the rare decay $B^0
  \to \jpsi\phi$}}, }{}\href{https://doi.org/10.1088/1674-1137/abdf40}{Chin.\
  Phys.\  \textbf{C45} (2021) 043001},
  \href{http://arxiv.org/abs/2011.06847}{{\normalfont\ttfamily
  arXiv:2011.06847}}\relax
\mciteBstWouldAddEndPuncttrue
\mciteSetBstMidEndSepPunct{\mcitedefaultmidpunct}
{\mcitedefaultendpunct}{\mcitedefaultseppunct}\relax
\EndOfBibitem
\bibitem{Verkerke:2003ir}
W.~Verkerke and D.~P. Kirkby, \ifthenelse{\boolean{articletitles}}{\emph{{The
  RooFit toolkit for data modeling}}, }{}eConf \textbf{C0303241} (2003)
  MOLT007, \href{http://arxiv.org/abs/physics/0306116}{{\normalfont\ttfamily
  arXiv:physics/0306116}}\relax
\mciteBstWouldAddEndPuncttrue
\mciteSetBstMidEndSepPunct{\mcitedefaultmidpunct}
{\mcitedefaultendpunct}{\mcitedefaultseppunct}\relax
\EndOfBibitem
\bibitem{LHCb:2021qbv}
LHCb collaboration, R.~Aaij {\em et~al.},
  \ifthenelse{\boolean{articletitles}}{\emph{{Precise measurement of
  the~$f_s/f_d$ ratio of fragmentation fractions and of $B^0_s$ decay branching
  fractions}}, }{}\href{https://doi.org/10.1103/PhysRevD.104.032005}{Phys.\
  Rev.\  \textbf{D104} (2021) 032005},
  \href{http://arxiv.org/abs/2103.06810}{{\normalfont\ttfamily
  arXiv:2103.06810}}\relax
\mciteBstWouldAddEndPuncttrue
\mciteSetBstMidEndSepPunct{\mcitedefaultmidpunct}
{\mcitedefaultendpunct}{\mcitedefaultseppunct}\relax
\EndOfBibitem
\bibitem{efron1979}
B.~Efron, \ifthenelse{\boolean{articletitles}}{\emph{Bootstrap methods: Another
  look at the jackknife},
  }{}\href{https://doi.org/10.1214/aos/1176344552}{Ann.\ Statist.\  \textbf{7}
  (1979) 1}\relax
\mciteBstWouldAddEndPuncttrue
\mciteSetBstMidEndSepPunct{\mcitedefaultmidpunct}
{\mcitedefaultendpunct}{\mcitedefaultseppunct}\relax
\EndOfBibitem
\bibitem{Cowan:2010js}
G.~Cowan, K.~Cranmer, E.~Gross, and O.~Vitells,
  \ifthenelse{\boolean{articletitles}}{\emph{{Asymptotic formulae for
  likelihood-based tests of new physics}},
  }{}\href{https://doi.org/10.1140/epjc/s10052-011-1554-0}{Eur.\ Phys.\ J.\
  \textbf{C71} (2011) 1554},
  \href{http://arxiv.org/abs/1007.1727}{{\normalfont\ttfamily
  arXiv:1007.1727}}, [Erratum: Eur.Phys.J.C 73, 2501 (2013)]\relax
\mciteBstWouldAddEndPuncttrue
\mciteSetBstMidEndSepPunct{\mcitedefaultmidpunct}
{\mcitedefaultendpunct}{\mcitedefaultseppunct}\relax
\EndOfBibitem
\bibitem{Schott:2012zb}
RooStats team collaboration, G.~Schott,
  \ifthenelse{\boolean{articletitles}}{\emph{{RooStats for searches}}, }{} in
  {\em {PHYSTAT 2011}}, \href{https://doi.org/10.5170/CERN-2011-006.199}{
  (Geneva), 199--208, CERN, 2011},
  \href{http://arxiv.org/abs/1203.1547}{{\normalfont\ttfamily
  arXiv:1203.1547}}\relax
\mciteBstWouldAddEndPuncttrue
\mciteSetBstMidEndSepPunct{\mcitedefaultmidpunct}
{\mcitedefaultendpunct}{\mcitedefaultseppunct}\relax
\EndOfBibitem
\end{mcitethebibliography}
